\documentclass[twocolumn]{aastex63}

\usepackage{amsmath}

\shortauthors{Margalit \& Quataert}
\shorttitle{synchrotron transients with thermal electrons}
\graphicspath{{./}{figures/}}

\begin{document}

\title{Thermal Electrons in Mildly-relativistic Synchrotron Blast-waves}

\author[0000-0001-8405-2649]{Ben Margalit}
\altaffiliation{NASA Einstein Fellow}
\affiliation{Astronomy Department and Theoretical Astrophysics Center, University of California, Berkeley, Berkeley, CA 94720, USA}

\author[0000-0001-9185-5044]{Eliot Quataert}
\affiliation{Department of Astrophysical Sciences, Princeton University, Princeton, NJ 08544, USA}




\begin{abstract} 
Numerical models of collisionless shocks robustly predict an electron distribution comprised of both thermal and non-thermal electrons. Here, we explore in detail the effect of thermal electrons on the emergent synchrotron emission from sub-relativistic shocks. We present a complete `thermal + non-thermal' synchrotron model and derive properties of the resulting spectrum and light-curves. Using these results we delineate  the relative importance of thermal and non-thermal electrons for sub-relativistic shock-powered synchrotron transients.   We find that thermal electrons are naturally expected to contribute significantly to the peak emission if the shock velocity is $\gtrsim 0.2c$, but would be mostly undetectable in non-relativistic shocks. This helps explain the dichotomy between typical radio supernovae and the emerging class of `AT2018cow-like' events. The signpost of thermal electron synchrotron emission is a steep optically-thin spectral index and a $\nu^2$ optically-thick spectrum. These spectral features are also predicted to correlate with a steep post-peak light-curve decline rate, broadly consistent with observed AT2018cow-like events. We expect that thermal electrons may be observable in other contexts where mildly-relativistic shocks are present, and briefly estimate this effect for gamma-ray burst afterglows and binary neutron star mergers. Our model can be used to fit spectra and light-curves of events and accounts for both thermal and non-thermal electron populations with no additional physical degrees of freedom.
\end{abstract}



\section{Introduction}
\label{sec:intro}

Synchrotron emission from relativistic electrons energized in astrophysical shock waves is observed in a wide variety of astrophysical sources.   It is typically assumed that strong collisionless shocks accelerate a {\it non-thermal} power-law distribution of relativistic electrons via diffusive-shock (first-order Fermi) acceleration \citep{Bell78,Blandford&Ostriker78,BlandfordEichler87}. This idea is directly supported both theoretically---by first-principles particle-in-cell (PIC) simulations \citep[e.g.][]{Spitkovsky08,Sironi&Spitkovsky09,Sironi&Spitkovsky11,Park+15}---and observationally---via power-law synchrotron emission that is observed ubiquitously in astrophysical sources.
Radio emission from non-relativistic shocks ($\beta_{\rm sh} \ll 1$; where $\beta_{\rm sh} c$ is the shock velocity) in interacting supernovae (SNe) has long been interpreted within a framework of non-thermal electron acceleration and shock amplified magnetic fields \citep[e.g.][]{Chevalier82,Weiler+02}. 
It is also understood that synchrotron self-absorption (SSA) plays a crucial role in such events \citep{Chevalier98}.
These ideas have also been successfully applied to ultra-relativistic explosions ($\Gamma_{\rm sh} \gg 1$, where $\Gamma_{\rm sh} = 1/\sqrt{1-\beta_{\rm sh}^2}$ is the shock Lorentz factor). Multi-wavelength observations of gamma-ray burst (GRB) afterglows
are well-interpreted as synchrotron emission produced by a non-thermal power-law distribution of relativistic electrons \citep{Sari+98}.

Recently, a new and enigmatic class of transients whose prototype is AT2018cow has been discovered \citep{Prentice+18,Margutti+19,Ho+19,Ho+20,Coppejans+20,Perley+21,Ho+21b,Bright+21}.
These events, discovered via optical transient surveys, are characterized by fast evolving bright optical emission, peculiar X-ray properties, and unusually luminous radio and millimeter (mm) emission.
There is currently no consensus model that easily explains all aspects of these events, but it is typically thought that circumstellar interaction must play an important role.
In particular, the radio--mm data is usually interpreted within a SSA framework similar to radio SNe models \citep{Chevalier98}. The shock velocities inferred from such modelling place these events in an unusual mildly-relativistic regime, with $0.1 \lesssim \beta_{\rm sh} \lesssim 0.5$.

\cite{Ho+21b} presented a detailed analysis of the radio and mm observations of AT2020xnd (see also \citealt{Bright+21}). The well sampled mm spectral-energy distribution (SED) of this event showed an unusually steep optically-thin spectral index $\alpha \approx -2$ (where $F_\nu \propto \nu^{\alpha}$) that is difficult to explain within the standard model of shock-powered synchrotron emission. Steep spectra were also observed for AT2018cow and CSS161010 at early epochs, potentially implicating a common physical mechanism.
This led \cite{Ho+21b} to propose that the observed steep spectra were due to synchrotron emission from a {\it thermal} population of electrons.
Using this model, \cite{Ho+21b} showed that the SEDs could be well-fit
if the peak (SSA) frequency is $\sim\mathcal{O}(100)$ times larger than the frequency 
at which `typical' thermal electrons emit.
For such parameters, the model could simultaneously explain both the steep optically-thin and shallow optically-thick slopes of observed events (see e.g. Fig.~11 in \citealt{Ho+21b}).
The success of this modelling makes the thermal electron scenario compelling, but it also raises many new questions:
Why would thermal electrons be observed for AT2018cow-like events but not in radio SNe or GRB afterglows?
How is this seemingly-peculiar property related to the unusually-bright and prolonged mm emission in such events?
Is it fine-tuned that the inferred SSA frequency is $\sim\mathcal{O}(100)$ times above the characteristic frequency at which most thermal electrons emit?

Separate from the specific observational motivation given above, numerical models of shock acceleration predict that most of the shock energy resides in the thermal population and that the non-thermal tail only contains a small fraction of the total post-shock energy \citep[e.g.][]{Park+15,Crumley+19}. Why then, should the non-thermal particles dominate the observed emission?   And are there physical conditions under which the thermal population is more observationally significant than has typically been assumed?

Here we address these questions by considering in greater detail the effect of thermal electrons on observed properties of synchrotron emission from sub-relativistic blast-waves.
The problem of synchrotron emission from a combined thermal + non-thermal electron distribution has previously been studied for ultra-relativistic shocks in GRBs \citep{Eichler&Waxman05,Giannios&Spitkovsky09,Ressler&Laskar17,Warren+18}, and separately in the context of hot accretion flows \citep[e.g.][]{Ozel+00}.
Here we study the case of sub-relativistic shocks ($\Gamma_{\rm sh} \beta_{\rm sh} < 1$) relevant to AT2018cow-like events, radio SNe, radio flares from binary neutron star (BNS) mergers, and GRBs at late times. We show that the shock velocity is the most important parameter that governs whether thermal electrons have an appreciable impact, and in particular that thermal electrons are naturally expected to dominate peak emission for mildly-relativistic shocks similar to those inferred for AT2018cow-like events.

This paper is organized as follows:
we begin in \S\ref{sec:model} by presenting the basic formalism and our model assumptions. This applies the results of \cite{Mahadevan+96} to the astrophysical setting of sub-relativistic shocks (analogous to \citealt{Ozel+00} who applied these results to hot accretion flows). We extend these results by considering the effects of synchrotron self-absorption and the fast-cooling regime for thermal electron synchrotron emission (\S\ref{sec:fast_cooling}). In \S\ref{sec:SED} we discuss the synchrotron spectrum resulting from this model (Fig.~\ref{fig:SEDs}) and derive expressions for relevant break frequencies in the problem (\ref{sec:break_frequencies}).
We subsequently present the landscape of sub-relativistic shock-powered synchrotron transients, illustrating a qualitative dichotomy between non-relativistic and mildly relativistic events (\S\ref{sec:phase_space}). We provide a brief discussion of the expected light-curves and temporal evolution in \S\ref{sec:time_evolution}, and conclude by discussing broader implications of our results (\S\ref{sec:discussion}).

\section{The Thermal + non-Thermal Model}
\label{sec:model}

We consider a sub-relativistic strong shock propagating with velocity $\beta_{\rm sh} c$ into a medium of density $n$.  For simplicity, we assume a constant shock compression ratio of $4$, formally valid  if the total post-shock energy density is dominated by non-relativistic particles.\footnote{Technically, in the parameter regime of interest to us, the shock compression ratio will depend mildly on the shock velocity, which sets whether the electrons are relativistic, and on the proton to electron temperature ratio, which sets how much of the post-shock thermal energy resides in the non-relativistic population.}  The downstream electron number density is then $n_e = 4\mu_e n$, where $\mu_e \simeq 1.18$ for Solar composition. 
PIC simulations show that, even when a non-thermal power-law tail of electrons is accelerated at the shock front, the majority of post-shock electrons do not participate in diffusive shock acceleration and instead occupy a quasi-thermal distribution.
Denoting the dimensionless temperature of these electrons as $\Theta \equiv k_B T_e / m_e c^2$, the post-shock electron energy density is
$u_e = a\left(\Theta\right) n_e \Theta m_e c^2$, where
\begin{equation}
    a\left(\Theta\right) \approx \frac{6+15\Theta}{4+5\Theta}
\end{equation}
is an approximation that is good to within $\sim 2\%$ \citep{Gammie&Popham98,Ozel+00}.
Assuming that ions govern the shock jump conditions (so that the effective adiabatic index is $5/3$ regardless of whether or not thermal electrons are relativistic), the total post-shock thermal energy density is 
$u = (9/8) n \mu m_p (\beta_{\rm sh}c)^2$, where $\mu \simeq 0.62$ for Solar composition.
If the electron ``thermalization efficiency'' is $\epsilon_T \lesssim 1$ such that $u_e = \epsilon_T u$ \citep{Margalit+21} then the post-shock electron temperature is,
\begin{align}
\label{eq:Theta}
    \Theta\left(\beta_{\rm sh}\right) 
    &\approx 
    \frac{ 5\Theta_0 - 6 + \sqrt{ 25\Theta_0^2 + 180\Theta_0 + 36 } }{30}
    \nonumber \\
    &\approx 
    \begin{cases}
    \frac{1}{3}\Theta_0 &,\, \Theta_0 \gg 1
    \\
    \frac{2}{3}\Theta_0 &,\, \Theta_0 \ll 1
    \end{cases}
\end{align}
where
\begin{equation}
\label{eq:Theta_0}
    \Theta_0 = \epsilon_T \frac{9\mu m_p}{32\mu_e m_e} \beta_{\rm sh}^2
    \approx 2.7 \epsilon_T \beta_{-1}^2
\end{equation}
and $\beta_{-1} \equiv \beta_{\rm sh}/0.1$.
The thermal electron population occupies a Maxwell-J\"{u}ttner distribution 
\begin{equation}
\label{eq:dn_dgamma_th}
    \left(\frac{\partial n}{\partial \gamma}\right)_{\rm th}
    = n_e f\left(\Theta\right) \sqrt{1-\gamma^{-2}} \frac{\gamma^2}{2 \Theta^3} e^{-\gamma/\Theta}
\end{equation}
where $\gamma$ is the electron Lorentz factor
and
\begin{equation}
    f(\Theta) \equiv 2\Theta^2/K_2(1/\Theta)
\end{equation}
is a correction-term that is only relevant in the non-relativistic regime ($f \approx 1$ for $\Theta \gtrsim 1$; $f \gg 1$ for $\Theta \ll 1$).

We additionally consider a non-thermal electron population which we model as a power-law distribution $\propto \gamma^{-p}$ (where $p>2$) that is terminated at some minimal Lorentz factor $\gamma_m$. We choose
\begin{equation}
\label{eq:gamma_m}
    \gamma_m(\Theta) = \langle \gamma \rangle_{\rm th} = 1+a(\Theta)\Theta
\end{equation}
equal to the mean Lorentz factor of thermal electrons $\langle \gamma \rangle_{\rm th}$, such that $\gamma_m \approx 3 \Theta$ ($\approx 1$) for $\Theta \gg 1$ ($\Theta \ll 1$). This choice is somewhat ad-hoc, but motivated by the fact that only supra-thermal electrons are capable of undergoing diffusive shock acceleration (the so-called `injection problem'; see e.g. \citealt{BlandfordEichler87}). 
Note that the exact value of $\gamma_m$ does not affect our results and is in any case degenerate with the assumed energy in the non-thermal tail.  The non-thermal electron distribution is then fully specified with only one additional parameter. We choose this to be $\delta$---the ratio of energy in the non-thermal electron population to that of thermal electrons (in terms of conventional $\epsilon_e$ notation, $\delta \equiv \epsilon_e/\epsilon_T$).
The non-thermal power-law distribution is therefore
\begin{equation}
    \left(\frac{\partial n}{\partial \gamma}\right)_{\rm pl}
    =
    n_e g(\Theta) \delta \frac{p-2}{3\Theta} \left(\frac{\gamma}{3\Theta}\right)^{-p}
\end{equation}
where 
\begin{equation}
    g(\Theta) \equiv \frac{(p-1) \gamma_m(\Theta) }{ (p-1)\gamma_m(\Theta)-p+2 } \left[\frac{\gamma_m(\Theta)}{3\Theta}\right]^{p-1}
\end{equation}
is a correction factor that is only important in the non-relativistic regime ($g \approx 1$ for $\Theta \gg 1$). 

Finally, we assume that plasma instabilities amplify magnetic fields in the downstream region with ``efficiency'' $\epsilon_B$, such that $B^2/8\pi = \epsilon_B u$. This implies
\begin{equation}
\label{eq:B}
    B = \sqrt{9\pi \epsilon_B n \mu m_p \beta_{\rm sh}^2 c^2}
    \approx 1.6 \,{\rm G}\,\epsilon_{B,-1}^{1/2} n_5^{1/2} \beta_{-1}
\end{equation}
where $n_5 = n/10^5\,{\rm cm}^{-3}$ is the upstream density, and $\epsilon_{B,-1} = \epsilon_B/0.1$.

The angle-averaged (assuming isotropic pitch angle distribution) synchrotron emissivity of the thermal electron population is 
\begin{equation}
\label{eq:jnu_th}
    j_{\nu,{\rm th}} = \frac{\sqrt{3} e^3 n_e B}{8\pi m_e c^2} f(\Theta) x I^\prime(x)
\end{equation}
where
$x \equiv {\nu}/{\nu_\Theta}$ and
\begin{equation}
\label{eq:nu_Theta}
    \nu_\Theta = \frac{3\Theta^2 e B}{4\pi m_e c}
    \underset{\Theta \gtrsim 1}{\approx} 
    5.6\,{\rm MHz}\,\epsilon_{B,-1}^{1/2} n_5^{1/2} \beta_{-1}^5 \epsilon_T^2
\end{equation}
is a scaling frequency that corresponds to the synchrotron frequency of electrons near the thermal peak when $\Theta \gtrsim 1$. 
Similarly, the absorption coefficient of thermal electrons is given by
\begin{equation}
\label{eq:alphanu_th}
    \alpha_{\nu,{\rm th}} = \frac{\pi e n_e}{3^\frac{3}{2} \Theta^5 B} f(\Theta) x^{-1} I^\prime(x)
    .
\end{equation}
In eqs.~(\ref{eq:jnu_th},\ref{eq:alphanu_th}) above, the function
\begin{equation}
\label{eq:I_of_x}
    I^\prime(x) \approx \frac{4.0505}{x^{1/6}} \left( 1 + \frac{0.40}{x^{1/4}} + \frac{0.5316}{x^{1/2}} \right) e^{-1.8899 x^{1/3}}
\end{equation}
is an approximation of the frequency dependence when $\Theta \gg 1$ \citep{Mahadevan+96}.
Although \cite{Mahadevan+96} provide different fitting coefficients as a function of $\Theta \lesssim 1$, we find it unnecessary to include these corrections here. As we later show, only high frequencies $x \gtrsim \mathcal{O}(10^2)$ are generally of interest (especially when $\Theta \lesssim 1$). In this regime $x \gg 1$ and eq.~(\ref{eq:I_of_x}) is exact \citep{Petrosian81}.

The pitch-angle averaged synchrotron emissivity of non-thermal electrons at frequencies $x \gg (\gamma_m/\Theta)^2$ is\footnote{
At lower frequencies the emissivity and absorption coefficient are affected by the power-law termination at $\gamma_m$, such that asymptotically $j_{\nu,{\rm pl}} \propto x^{1/3}$ and $\alpha_{\nu,{\rm pl}} \propto x^{-5/3}$. We include this in our numeric calculations for completeness, but remark that this has no affect on any of our results, and can therefore be neglected.
}
\begin{equation}
\label{eq:jnu_pl}
    j_{\nu,{\rm pl}} = C_j \frac{e^3 n_e B \delta}{m_e c^2} g(\Theta) x^{-\frac{p-1}{2}} 
\end{equation}
where
\begin{equation}
    C_j = \frac{ 3^\frac{2p-1}{2} (p-2) \Gamma\left(\frac{p+5}{4}\right) \Gamma\left(\frac{3p+19}{12}\right) \Gamma\left(\frac{3p-1}{12}\right) }{ 2^\frac{7-p}{2} \pi^\frac{1}{2} (p+1) \Gamma\left(\frac{p+7}{4}\right) }
    .
\end{equation}
The non-thermal absorption coefficient is similarly
\begin{equation}
\label{eq:alphanu_pl}
    \alpha_{\nu,{\rm pl}} = C_\alpha \frac{e n_e \delta}{\Theta^5 B} g(\Theta) x^{-\frac{p+4}{2}}
\end{equation}
with 
\begin{equation}
    C_\alpha = \frac{ 2^\frac{p}{2} \pi^\frac{3}{2} (p-2) \Gamma\left(\frac{p+6}{4}\right) \Gamma\left(\frac{3p+2}{12}\right) \Gamma\left(\frac{3p+22}{12}\right) }{ 3^\frac{5-2p}{2} \Gamma\left(\frac{p+8}{4}\right) }
    .
\end{equation}
Note that pitch-angle averaging has often been neglected in many previous applications of non-thermal synchrotron emission \citep[e.g.][]{Chevalier98}. We include this order-unity correction here because it is expected in the standard scenario where magnetic fields are turbulently amplified, and because thermal electrons are more appreciably affected by such averaging \citep{Mahadevan+96}.

\subsection{Fast Cooling}
\label{sec:fast_cooling}

Observed emission depends on line-of-sight integrals of the emissivity and absorption coefficients. 
Synchrotron emitting electrons whose Lorentz factor exceeds
\begin{equation}
\label{eq:gamma_cool}
    \gamma_{\rm cool} = \frac{6\pi m_e c}{\sigma_T B^2 t}
    \approx 34 \epsilon_{B,-1}^{-1} n_5^{-1} \beta_{-1}^{-2} t_{100}^{-1}
\end{equation}
will radiate most of their energy over a timescale that is short compared to the dynamical time, $t = 100\,{\rm d}\,t_{100}$. Such fast-cooling electrons would only reside within a fractional depth $\sim (\gamma / \gamma_{\rm cool})^{-1} \ll 1$ behind the shock front, so that the effective line-of-sight averaged distribution function is 
$\langle \partial n / \partial \gamma \rangle  \sim (\partial n / \partial \gamma) \min( 1, \gamma_{\rm cool}/\gamma )$.
The emissivity and absorption coefficient are proportional to this distribution function and therefore similarly affected, introducing a frequency-dependent correction to eqs.~(\ref{eq:jnu_th},\ref{eq:alphanu_th},\ref{eq:jnu_pl},\ref{eq:alphanu_th}) above.

For power-law electrons, emission/absorption at frequency $x$ is contributed predominantly by electrons of Lorentz factor $\gamma/\Theta \sim x^{1/2}$. This implies the standard fast-cooling correction
\begin{equation}
\label{eq:cooling_correction_pl}
    \begin{Bmatrix} \left\langle j_{\nu,{\rm pl}} \right\rangle \\ \left\langle \alpha_{\nu,{\rm pl}} \right\rangle \end{Bmatrix} 
    \sim
    \begin{Bmatrix} j_{\nu,{\rm pl}} \\ \alpha_{\nu,{\rm pl}} \end{Bmatrix} 
    \min \left[ 1, \left(\frac{x}{x_{\rm cool,pl}}\right)^{-1/2} \right]
\end{equation}
where $x_{\rm cool,pl} \equiv (\gamma_{\rm cool}/\Theta)^2$.
For thermal electrons however, this no longer applies. At frequencies $x \gg 1$ (of main interest here), emission/absorption samples the high-frequency tail of comparatively lower-energy electrons (which vastly outnumber electrons at higher Lorentz factors). 
In this regime, the synchrotron frequency $x$ is instead related to electrons whose characteristic Lorentz factor is $\gamma/\Theta \sim (2x)^{1/3}$,\footnote{
This can be shown by examining the synchrotron integral of the thermal population, $\propto \int z^2 e^{-z} F(x/z^2) dz$, where $z \equiv \gamma/\Theta$. At high frequencies $x \gg 1$ this is $\propto \int z e^{-(z+x/z^2)} dz$, and the integral is dominated by electrons with Lorentz factors near $z = (2x)^{1/3}$, at which the function in the exponent attains a minimum (this is equivalent to the method of steepest descent approach that was used in deriving eq.~\ref{eq:I_of_x}; see \citealt{Petrosian81}).
}
and the cooling-corrected emissivity and absorption coefficients are
\begin{equation}
\label{eq:cooling_correction_th}
    \begin{Bmatrix} \left\langle j_{\nu,{\rm th}} \right\rangle \\ \left\langle \alpha_{\nu,{\rm th}} \right\rangle \end{Bmatrix} 
    \underset{x \gg 1}{\sim}
    \begin{Bmatrix} j_{\nu,{\rm th}} \\ \alpha_{\nu,{\rm th}} \end{Bmatrix} 
    \min \left[ 1, \left(\frac{x}{x_{\rm cool,th}}\right)^{-1/3} \right]
\end{equation}
where $x_{\rm cool,th} \equiv (\gamma_{\rm cool}/\Theta)^3/2$.

The total thermal + non-thermal emissivity is simply $\langle j_\nu \rangle = \langle j_{\nu,{\rm th}} \rangle + \langle j_{\nu,{\rm pl}} \rangle$, and the combined absorption coefficient is similarly additive, so that $\langle \alpha_\nu \rangle = \langle \alpha_{\nu,{\rm th}} \rangle + \langle \alpha_{\nu,{\rm pl}} \rangle$.
The emergent specific luminosity is then
\begin{equation}
\label{eq:Lnu}
    L_\nu = 4\pi^2 R^2 \frac{\langle j_\nu \rangle}{\langle \alpha_\nu \rangle} \left( 1 - e^{-\langle \alpha_\nu \rangle R} \right) ,
\end{equation}
where $R$ is the characteristic size of the emitting region, and the effective absorption and emission coefficients are given by eqs.~(\ref{eq:jnu_th},\ref{eq:alphanu_th},\ref{eq:jnu_pl},\ref{eq:alphanu_pl},\ref{eq:cooling_correction_pl},\ref{eq:cooling_correction_th}) above.

\section{Spectrum}
\label{sec:SED}

The synchrotron spectrum that results from the `thermal + non-thermal' model (eq.~\ref{eq:Lnu}) typically peaks at the SSA frequency $\nu_{\rm a}$. At frequencies $\nu < \nu_{\rm a}$ emission is self-absorbed and the spectrum rises as a function frequency, whereas above it emission is optically-thin and the spectral luminosity decreases with frequency.
There are two distinct regimes that are of particular interest: (i) emission near the SSA frequency is dominated by thermal electrons; or (ii) emission near this frequency is instead dominated by the power-law electron distribution.

\begin{figure*}
    \includegraphics[width=0.5\textwidth]{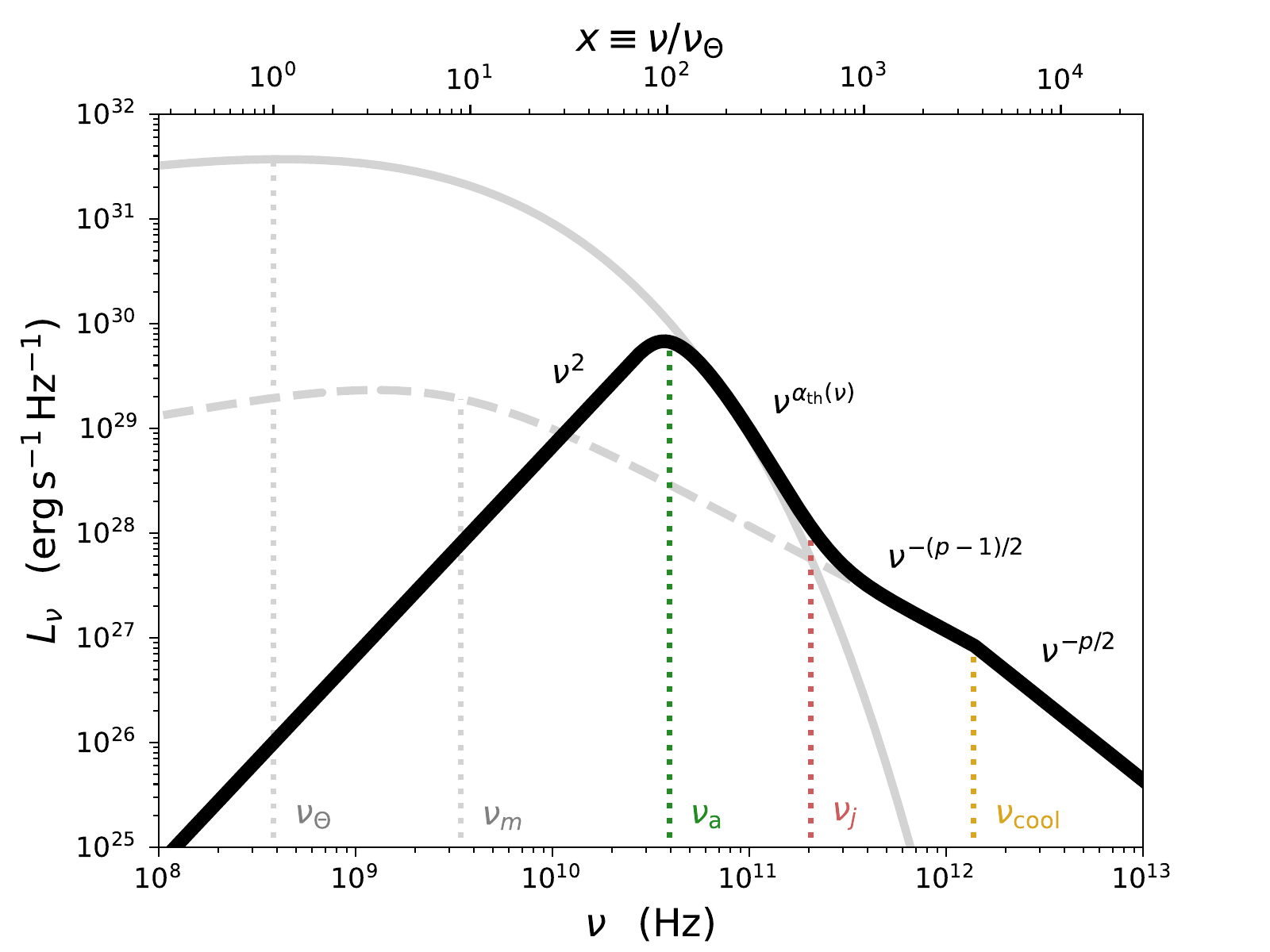}
    \includegraphics[width=0.5\textwidth]{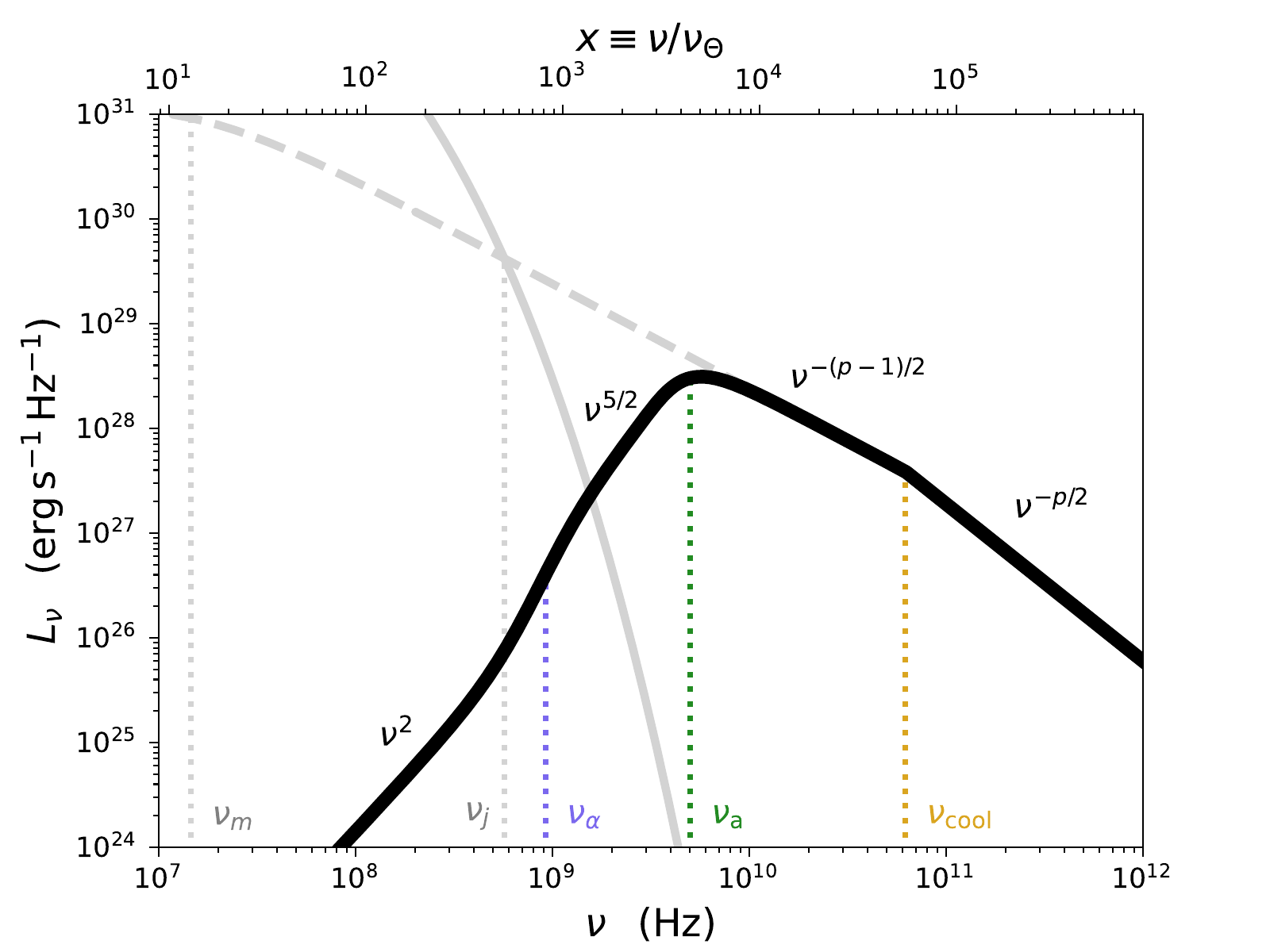}
    \caption{Example SEDs. 
    {\it Left:} A ``thermal spectrum'', where peak emission is dominated by thermal electrons ($\nu_{\rm a} < \nu_j$). Solid (dashed) light-grey curves show the optically-thin contribution of thermal (power-law) electrons to the SED, while the black curve shows the combined emergent spectral luminosity, including synchrotron self-absorption (eq.~\ref{eq:Lnu}). Vertical dotted curves show characteristic break frequencies
    (see Table~\ref{tab:frequencies}): 
    the thermal synchrotron frequency $\nu_\Theta$ (eq.~\ref{eq:nu_Theta}); the frequency $\nu_m$ corresponding to the minimal Lorentz factor of power-law electrons; the synchrotron self-absorption frequency $\nu_{\rm a}$ (eqs.~\ref{eq:xa_th_slowcooling},\ref{eq:xa_th_fastcooling},\ref{eq:xa_pl}); the frequency $\nu_j$ ($\nu_\alpha$) at which emission (absorption) transitions from being dominated by thermal electrons to power-law electrons (eqs.~\ref{eq:xj},\ref{eq:x_alpha}); and the synchrotron cooling frequency $\nu_{\rm cool}$ (eqs.~\ref{eq:nu_cool_th},\ref{eq:alphanu_pl}).
    The spectral slope (stated above each segment) can be especially steep in the optically-thin thermal regime, $\nu_{\rm a} < \nu < \nu_j$ (eq.~\ref{eq:th_spectral_index}).
    {\it Right:} Same as left panel, but for a ``non-thermal spectrum'' where peak emission is dominated by the power-law electron distribution ($\nu_{\rm a} > \nu_j$). The SED follows the standard non-thermal spectrum, except at very low frequencies $< \nu_\alpha$ (eq.~\ref{eq:x_alpha}) where the SSA spectrum softens. The thermal electron population would be mostly unobservable in this regime, even though it is energetically (and by number) dominant.
    Both panels show cases where $\nu_{\rm cool}$ falls above other relevant frequencies, but alternative orderings may be possible and are fully accounted for in \S\ref{sec:break_frequencies}).
    The left panel is calculated using $\beta_{\rm sh}=0.45$, $n=10^3\,{\rm cm}^{-3}$, and $t=25\,{\rm d}$. The right panel is for $\beta_{\rm sh}=0.1$, $n=10^4\,{\rm cm}^{-3}$, and $t=200\,{\rm d}$. In both cases we assume $\delta=0.01$, $p=3$, $\epsilon_B=0.1$, and $\epsilon_T=1$.
    }
    \label{fig:SEDs}
\end{figure*}

Figure~\ref{fig:SEDs} shows representative SEDs in these two cases.
Solid (dashed) light-grey curves show the optically-thin thermal (power-law) electron emission, while solid black curves show the combined spectrum including self-absorption (eq.~\ref{eq:Lnu}). 
Vertical dotted curves show relevant break frequencies. These are listed in Table~\ref{tab:frequencies} and discussed in greater detail in \S\ref{sec:break_frequencies}.
The left panel shows a ``thermal spectrum'' where peak emission is governed by thermal electrons.
At low frequencies $\nu < \nu_{\rm a}$ the SED follows the Rayleigh-Jeans limit $\propto \nu^2$. This is shallower than the canonical $\nu^{5/2}$ SED of optically-thick power-law synchrotron emission \citep{Rybicki&Lightman79}.
At frequencies slightly above peak the SED follows the optically-thin thermal emissivity and the spectral slope can be extremely steep. The spectrum does not follow a power-law-form in this regime, however we can characterize the slope steepness via the
frequency-dependent spectral index
$\alpha_{\rm th} \equiv {d\ln \langle j_{\nu,{\rm th}} \rangle}/{d\ln \nu}$,
\begin{equation}
\label{eq:th_spectral_index}
    \alpha_{\rm th}(\nu) 
    \underset{\nu \gg \nu_\Theta}{\approx}
    \begin{cases}
    \frac{5}{6} - \left(\frac{\nu}{4 \nu_\Theta}\right)^{1/3}
    &, \nu < \nu_{\rm cool}
    \\ 
    \frac{1}{2} - \left(\frac{\nu}{4 \nu_\Theta}\right)^{1/3}
    &, \nu > \nu_{\rm cool}
    \end{cases}
    .
\end{equation}
The expression above applies in the typical setting where frequencies of interest are $\gg \nu_\Theta$ (eq.~\ref{eq:nu_Theta}), and the two cases (whose spectral slope differs by $1/3$) depend on whether the observing frequency is below or above the fast-cooling break frequency $\nu_{\rm cool}$.

The spectral index implied by eq.~(\ref{eq:th_spectral_index}) becomes increasingly steep at higher frequencies, and is a unique feature of the thermal electron model. However, above some frequency $\nu_j$, emission by power-law electrons will come to dominate the thermal-electron emission. This transition frequency (red dotted curve in Fig.~\ref{fig:SEDs}) depends primarily on the relative number of power-law and thermal electrons, which is $\propto \delta$ in our model. Lower values of $\delta$ imply a smaller fraction of power-law electrons and a higher transition frequency (in \S\ref{sec:break_frequencies} we provide approximate expressions for this dependence).

Fig.~\ref{fig:SEDs} illustrates the significance of $\nu_j$. At frequencies $\nu > \nu_j$ emission is governed by non-thermal electrons and the spectrum follows standard results for power-law synchrotron emission---the SED is $\propto \nu^{-(p-1)/2}$ ($\propto \nu^{-p/2}$) in the slow- (fast-) cooling optically-thin regimes.
If a given event is only observed at frequencies $\nu > \nu_j$ then the thermal-electron contribution would go undetected and this would be indistinguishable from a purely non-thermal electron model.

This is further illustrated by the right-hand panel of Fig.~\ref{fig:SEDs}, which shows a ``non-thermal spectrum'' where the peak (SSA) frequency is $\nu_{\rm a} > \nu_j$. In this case peak emission is dominated by the power-law electron distribution, the usual $\nu^{5/2}$ SSA optically-thick spectrum applies below peak, and the entire optically-thin SED follows the standard power-law spectrum. Thermal electrons---though present and energetically dominant in this model--- would not affect the observed emission except at very low frequencies $\nu < \nu_\alpha \sim \nu_j$ where the optically-thick spectrum is expected to soften (at $\nu < \nu_\alpha$ the optical depth becomes dominated by thermal electrons). These frequencies are usually observationally inaccessible so that our `thermal + non-thermal' model would be indistinguishable from purely non-thermal synchrotron models that are typically used to model observations.
We also note that the low frequency spectrum is sensitive to geometric effects (related to the spatial distribution of emitting electrons) and can be susceptible to scintillation, further complicating potential identification of a break frequency at $\nu_\alpha$.

In the following subsection we discuss the various break frequencies shown in Fig.~\ref{fig:SEDs} in greater detail. Readers interested primarily in our main results may wish to skip forward to \S\ref{sec:phase_space}, while those interested in understanding the origin of different regions and scaling with physical parameters are welcome to continue to \S\ref{sec:break_frequencies}.

\subsection{Estimates of Break Frequencies}
\label{sec:break_frequencies}

\begin{deluxetable*}{lll}[]
\tablecaption{Key Frequencies and their definitions. 
See \S\ref{sec:SED} for further details.
\label{tab:frequencies}}
\tablewidth{0pt} 
\tablehead{ \colhead{notation$^{a}$} & \colhead{equation} & \colhead{definition}}
\tabletypesize{\small} 
\startdata 
$\nu_\Theta$ & eq.~(\ref{eq:nu_Theta}) & characteristic synchrotron frequency of thermal electrons\\
$\nu_m$ & eq.~(\ref{eq:gamma_m}) & characteristic synchrotron frequency of power-law electrons, $\nu_m = \left(\gamma_m/\Theta\right)^2 \nu_\Theta$\\
$\nu_{\rm cool}$ & eqs.~(\ref{eq:nu_cool_pl},\ref{eq:nu_cool_th}) & synchrotron cooling frequency\\
$\nu_{\rm a}$ & eq.~(\ref{eq:nu_a}) & synchrotron self-absorption (SSA) frequency, $\langle \alpha_{\nu} \rangle R = 1$ at $\nu_{\rm a}$\\
$\nu_j$ & eq.~(\ref{eq:xj}) & frequency above which power-law $>$ thermal emissivity, $\langle j_{\nu,{\rm th}} \rangle = \langle j_{\nu,{\rm pl}} \rangle$ at $\nu_j$\\
$\nu_\alpha$ & eq.(~\ref{eq:x_alpha}) & frequency above which power-law $>$ thermal absorption, $\langle \alpha_{\nu,{\rm th}} \rangle = \langle \alpha_{\nu,{\rm pl}} \rangle$ at $\nu_\alpha$\\
\enddata 
\centering
\tablenotetext{a}{Note that normalized frequency $x \equiv \nu/\nu_\Theta$ is used interchangeably with $\nu$ throughout the text.}
\end{deluxetable*} 

As illustrated by Fig.~\ref{fig:SEDs},
the resulting SED of the thermal + non-thermal model depends on several characteristic frequencies.
The first is the `thermal' frequency $\nu_\Theta$ given by eq.~(\ref{eq:nu_Theta}). Many other relevant frequencies scale in some well-determined way with $\nu_\Theta$.
For example, the frequency $\nu_m$ that corresponds to the minimum Lorentz factor of power-law electrons is simply $\nu_m = (\gamma_m/\Theta)^2 \nu_\Theta$ (and is $\nu_m \approx 9 \nu_\Theta$ for $\Theta \gtrsim 1$).

The synchrotron cooling frequency $\nu_{\rm cool}$ can also affect the observed SED. As discussed in \S\ref{sec:fast_cooling}, this frequency is 
related to $\gamma_{\rm cool}$ (eq.~\ref{eq:gamma_cool}) as
\begin{align}
\label{eq:nu_cool_pl}
    \nu_{\rm cool, pl} 
    &= \left(\frac{\gamma_{\rm cool}}{\Theta}\right)^2 \nu_\Theta
    = \frac{27 \pi m_e c e}{\sigma_T^2 B^3 t^2}
    \\ \nonumber
    &\approx 7.9\,{\rm GHz}\, \epsilon_{B,-1}^{-3/2} n_5^{-3/2} \beta_{-1}^{-3} t_{100}^{-2}
\end{align}
for power-law emitting electrons, and
\begin{align}
\label{eq:nu_cool_th}
    \nu_{\rm cool,th}
    &= 
    \frac{1}{2} \left(\frac{\gamma_{\rm cool}}{\Theta}\right)^3 \nu_\Theta
    = \frac{81 \pi^2 m_e^2 c^2 e}{\sigma_T^3 B^5 t^3 \Theta}
    \\ \nonumber
    &\underset{\Theta \gtrsim 1}{\approx} 150\,{\rm GHz}\, \epsilon_T^{-1} \epsilon_{B,-1}^{-5/2} n_5^{-5/2} \beta_{-1}^{-7} t_{100}^{-3}
\end{align}
for the thermal electron population.
The observed cooling break therefore depends on whether emission is dominated by power-law or thermal electrons.
This is governed by the frequency $\nu_j$ at which the thermal and power-law emissivities are equal, $\langle j_{\nu,{\rm th}} \rangle = \langle j_{\nu,{\rm pl}} \rangle$.
The transcendental equation for $\nu_j$ does not permit a closed form analytic solution, but is easily solvable numerically. 
In general, the solution depends on $\delta$, $p$, and $\Theta$, however the $\Theta$ dependence is suppressed for $\Theta \gtrsim 1$.
An accurate fitting function to the solution is given by
\begin{align}
\label{eq:xj}
    x_j(\Theta \gtrsim 1) \approx 40.94 - 49.97 \ln\left(\delta\right)
    + 12.51 \ln\left(\delta\right)^2
\end{align}
where $x_j \equiv \nu_j / \nu_\Theta$ following our standard notation.
This is accurate to within 3\% for $10^{-6} \leq \delta \leq 1/3$, $p=3$, and any $\Theta \gg 1$, but is also reasonably accurate for $\Theta \gtrsim 1$ or other values of $2.2 \leq p \leq 3.4$ (17\% accuracy).
For a fiducial $\delta = 0.01$, we find that $x_j \approx 540$.
An alternative approximation that is accurate to within 19\% between $10^{-5} \leq \delta \leq 0.1$ is given by $x_j \approx 150 \delta^{-0.25}$.

A related frequency $\nu_\alpha$ is defined by equating the absorption coefficients of the two populations such that $\langle \alpha_{\nu,{\rm th}} \rangle = \langle \alpha_{\nu,{\rm pl}} \rangle$ at $\nu_\alpha$.
This frequency is typically a factor $\lesssim 2$ greater than $\nu_j$, and we find that the approximation
\begin{equation}
\label{eq:x_alpha}
    x_\alpha(\Theta \gtrsim 1) \approx 5.221 x_j \ln\left(x_j\right)^{-0.6373}
\end{equation}
is accurate to within several percent throughout the parameter range considered above.

Finally, the SED peak is set by the SSA frequency $\nu_{\rm a}$. If the absorption coefficient is dominated by thermal electrons ($x < x_\alpha$) then the SSA frequency is determined by the condition $\langle \alpha_{\nu,{\rm th}} \rangle R = 1$ (eqs.~\ref{eq:alphanu_th},\ref{eq:cooling_correction_th}).
In the slow-cooling regime, this is governed by an optical-depth parameter
\begin{align}
\label{eq:tau_Theta}
    \tau_\Theta 
    \equiv \frac{\pi e n_e R}{3^\frac{3}{2} \Theta^5 B} f(\Theta)
    \underset{\Theta \gtrsim 1}{\approx} 3.6 \times 10^{12}\, \epsilon_T^{-5} \epsilon_{B,-1}^{-1/2} n_5^{1/2} \beta_{-1}^{-10} t_{100}
\end{align}
that describes the (thermal-contribution to the) optical depth at frequency $\approx \nu_\Theta$.
Note the extreme sensitivity of $\tau_\Theta$ to the shock velocity and the fact that $\tau_\Theta \gg 1$ for typical parameters. This implies that one would not expect to see ``bare''(unabsorbed) Maxwellian SEDs that peak at $\sim \nu_\Theta$. We elaborate on this in \S\ref{sec:phase_space}.

Using the optical-depth parameter $\tau_\Theta$ (eq.~\ref{eq:tau_Theta}), we find that the thermal SSA frequency is well approximated by the fitting function
\begin{equation}
\label{eq:xa_th_slowcooling}
    x_{\rm a,th}(<x_{\rm cool,th}) \approx 
    \left( \frac{3.434}{\ln\left(\tau_\Theta\right)} - \frac{4.762}{\ln\left(\tau_\Theta\right)^2} - 0.028 \right)^{-3}
\end{equation}
that is accurate to within 3\% over many orders of magnitude in optical depth, $30 \leq \tau_\Theta \leq 10^{12}$.
The SSA frequency varies between $x_{\rm a,th} \sim 10-10^3$ over this range of $\tau_\Theta$.
An alternative simpler approximation that is accurate to within 18\% for $5\times10^3 \leq \tau_\Theta \leq 2 \times 10^{11}$ 
is given by $x_{\rm a,th} \approx 7.1 \tau_\Theta^{0.2}$.
In the fast cooling regime $x_{\rm a,th} > x_{\rm cool, th}$ the fitting functions above are comparably accurate when transformed as
\begin{equation}
\label{eq:xa_th_fastcooling}
    x_{\rm a,th}(>x_{\rm cool, th})
    \approx
    0.68 
    \times
    {\rm (eq.~28);~} \tau_\Theta \to 
    \frac{\gamma_{\rm cool}}{\Theta}\tau_\Theta
    .
\end{equation}

If the absorption coefficient is instead dominated by power-law electrons then the SSA frequency is determined by the condition $\langle \alpha_{\nu,{\rm pl}} \rangle R = 1$. This results in the analytic solution
\begin{equation}
\label{eq:xa_pl}
    x_{\rm a, pl}
    = 
    \begin{cases}
    \left( \frac{C_\alpha e \delta n_e R g\left(\Theta\right)}{\Theta^5 B} \right)^\frac{2}{p+4}
    &,x_{\rm a, pl}<x_{\rm cool,pl}
    \\
    \left( \frac{C_\alpha e \delta n_e R g\left(\Theta\right) \gamma_{\rm cool}}{\Theta^6 B} \right)^\frac{2}{p+5}
    &,x_{\rm a, pl}>x_{\rm cool,pl}
    \end{cases}
    .
\end{equation}
Noting that $g(\Theta) \approx (p-1) (3\Theta)^{-(p-1)}$ when $\Theta \ll 1$ and $g(\Theta) \approx 1$ for $\Theta \gtrsim 1$, we can express the power-law-dominated SSA frequency as
\begin{align}
    \nu_{\rm a,pl}(<\nu_{\rm cool,pl})
    \underset{p=3}{\approx}
    &\epsilon_{B,-1}^{5/14} \delta_{-2}^{2/7} n_5^{9/14} t_{100}^{2/7}
    \\ \nonumber
    &\times
    \begin{cases}
    12\,{\rm GHz}\, \beta_{-1} 
    &, \Theta \ll 1
    \\
    17\,{\rm GHz}\, 
    \epsilon_T^{4/7}  \beta_{-1}^{15/7}
    &, \Theta \gtrsim 1
    \end{cases}
\end{align}
in the slow-cooling regime, and
\begin{align}
    \nu_{\rm a,pl}(>\nu_{\rm cool,pl})
    \underset{p=3}{\approx}
    &\epsilon_{B,-1}^{1/8} \delta_{-2}^{1/4} n_5^{3/8}
    \\ \nonumber
    &\times
    \begin{cases}
    34\,{\rm GHz}\, \beta_{-1}^{1/2}
    &, \Theta \ll 1
    \\
    16\,{\rm GHz}\, \epsilon_T^{1/2}  \beta_{-1}^{3/2}
    &, \Theta \gtrsim 1
    \end{cases}
\end{align}
in the fast-cooling case, and we have chosen a fiducial $p=3$ for the estimates above.
In total, the SSA frequency is related to eqs.~(\ref{eq:xa_th_slowcooling},\ref{eq:xa_th_fastcooling},\ref{eq:xa_pl}),
\begin{equation}
\label{eq:nu_a}
    \nu_{\rm a} = 
    \begin{cases}
    \nu_{\rm a, th} &,~ \nu_{\rm a} < \nu_\alpha
    \\
    \nu_{\rm a, pl} &,~ \nu_{\rm a} > \nu_\alpha
    \end{cases}
    .
\end{equation}
The different frequencies discussed in this section are also summarized in Table~\ref{tab:frequencies}.

\section{Phase-space of Transients}
\label{sec:phase_space}

In the previous section we showed that the ordering of characteristic frequencies (primarily the self-absorption frequency $\nu_{\rm a}$ and the frequency $\nu_j$ at which thermal and non-thermal electrons have comparable emissivity) determines whether thermal electrons contribute appreciably to observed emission. This is illustrated by the different spectra in the left and right hand panels of Fig.~\ref{fig:SEDs}.
A natural question subsequently arises---{\it 
what type of shock-powered transients might be expected to show signatures of a thermal electron distribution?
}

\begin{figure*}
    \centering
    \includegraphics[width=0.9\textwidth]{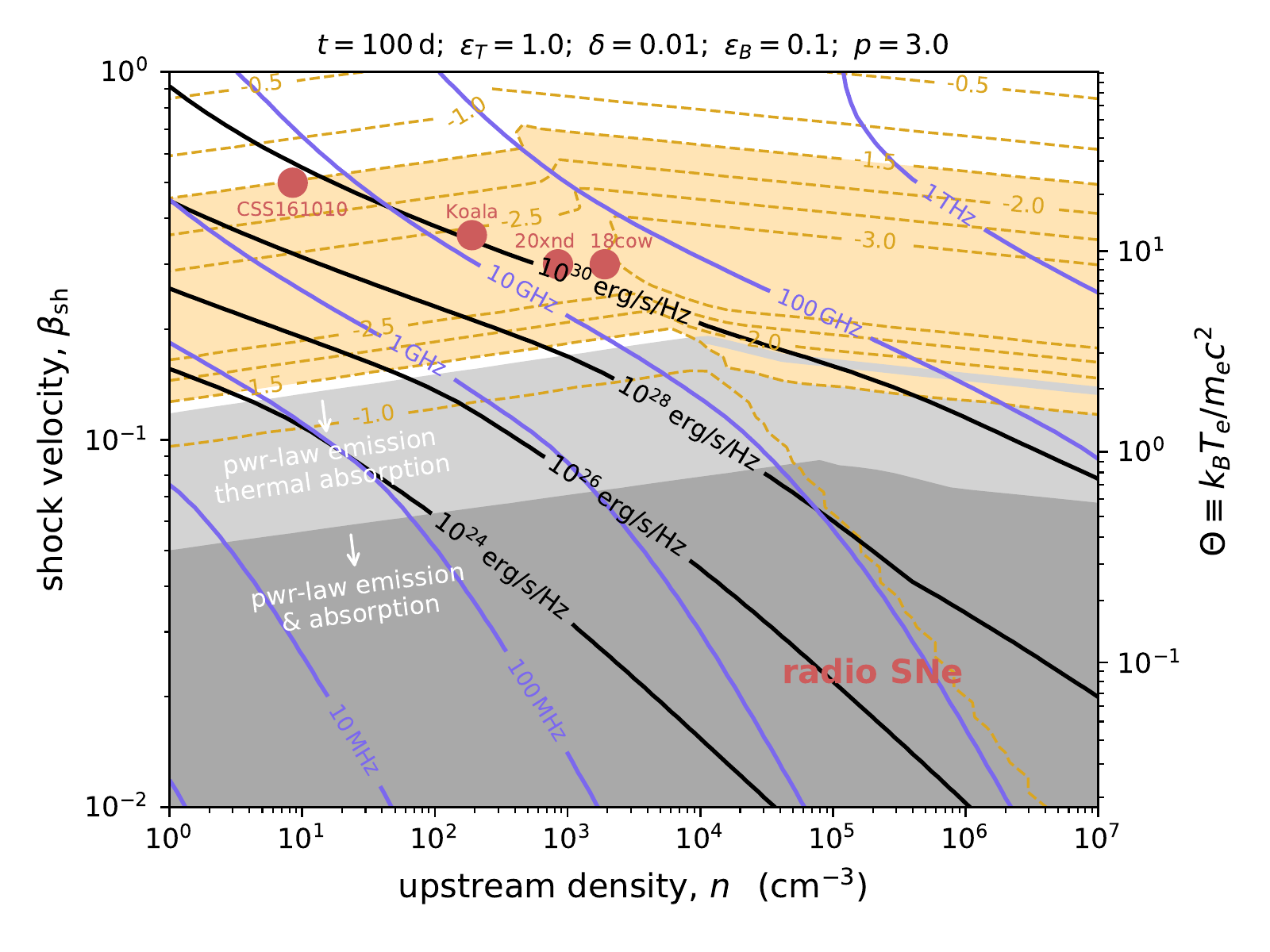}
    \caption{Parameter space of shock-powered sub-relativistic synchrotron transients, where both thermal and non-thermal (power-law) electrons are considered. At a given epoch, $t = 100\,{\rm d}$, blue curves show contours of frequency at which the SED peaks ($\approx \nu_{\rm a}$; set by SSA) as a function of shock velocity $\beta_{\rm sh} c$ and ambient density $n$. Black contours show the (peak) specific luminosity at this frequency.
    Yellow dashed contours show the spectral index slightly above this peak frequency (at $\nu = 2 \times$peak).
    Within the light (dark) grey shaded region emission (and absorption) are dominated by the power-law electron distribution and thermal electrons would not affect the observations (see Fig.~\ref{fig:SEDs}, right panel). Outside these regions emission near the peak (SSA) frequency is instead dominated by thermal electrons (see left panel of Fig.~\ref{fig:SEDs}). This may be observationally distinguishable from a purely power-law electron model by the unusually steep optically-thin spectral slope (yellow shaded region).
    The shock velocity is the most important parameter that governs whether a steep ``thermal'' or a standard ``non-thermal'' spectrum would be observable (eq.~\ref{eq:th_electron_condition}). The dichotomy between radio SNe and AT2018cow-like events can therefore be naturally understood as an artefact of the non-relativistic ($\beta_{\rm sh} \ll 1$) vs mildly-relativistic ($\beta_{\rm sh} \gtrsim 0.2$) velocities inferred for these events
    (see \S\ref{fig:phase_space}).
    }
    \label{fig:phase_space}
\end{figure*}

Figure~\ref{fig:phase_space} addresses this question by showing the parameter-space of sub-relativistic shock-powered synchrotron transients.
This phase space is determined by the upstream ambient density $n$, the shock velocity $\beta_{\rm sh}c$, and the size of the emitting region $R$. We relate the size to the shock velocity 
as $R = \beta_{\rm sh} c t$ such that $t$ is an effective dynamical time (fixed to $100\,{\rm d}$ in Fig.~\ref{fig:phase_space}). 
This corresponds to the true time-since-explosion only if the shock velocity is temporally constant. If the shock decelerates then this time parameter would be larger than the actual observing epoch (non-spherical geometry can also affect this).
The electron temperature and magnetic field are directly related to $\beta_{\rm sh}$, $n$ through eqs.~(\ref{eq:Theta},\ref{eq:Theta_0},\ref{eq:B}), and we 
adopt fiducial values $\epsilon_T=1$, $\epsilon_B = 0.1$, $\delta = 0.01$, and $p=3$.

Blue contours in Fig.~\ref{fig:phase_space} show the frequency at which the SED peaks. Throughout nearly the entire illustrated parameter space the thermal optical-depth is $\tau_\Theta \gg 1$ so the peak frequency is $\approx \nu_{\rm a}$ (eq.~\ref{eq:nu_a}) as set by SSA. Black contours show the peak specific luminosity at this frequency (eq.~\ref{eq:Lnu}).
Grey shaded regions show the parameter space in which optically-thin emission is set entirely by the power-law electron distribution. This is determined by the condition $\nu_j < \nu_{\rm a}$ (eqs.~\ref{eq:xj},\ref{eq:xa_pl}) that implies a spectrum similar to the right panel in Fig.~\ref{fig:SEDs}.
Within the light-grey region $\nu_\alpha \gtrsim \nu_{\rm a}/5$ and the presence of thermal electrons may still be discernible through their effect on the self-absorbed spectrum: between $\nu_\alpha < \nu < \nu_{\rm a}$ the SSA spectrum follows the canonical $\nu^{5/2}$ scaling of a power-law electron distribution, but at frequencies $\nu < \nu_\alpha$ this softens to a thermal SSA spectrum $\propto \nu^2$ (see right panel in Fig.~\ref{fig:SEDs}). In the dark shaded grey region $\nu_\alpha \ll \nu_{\rm a}$ so that this softening would occur well below the SED peak and would be more difficult to detect.

Finally, we also plot in Fig.~\ref{fig:phase_space} contours of the spectral index just above the SED peak (at frequency $\nu \approx 2\nu_{\rm a}$). For power-law electrons with our canonical $p=3$ this spectral index would be $-(p-1)/2 = -1$ in the slow-cooling regime and $-p/2 = -1.5$ in the fast-cooling case. 
The transition between fast and slow cooling regimes is apparent through the kink in the spectral-index contours.
Alternatively, if emission near $\nu_{\rm a}$ is dominated by the thermal electron population then the spectral index can be significantly steeper (eq.~\ref{eq:th_spectral_index}; left panel, Fig.~\ref{fig:SEDs}). We highlight this with the yellow shaded area in Fig.~\ref{fig:phase_space}, which shows regions where the spectral index is steeper than would be expected for purely power-law electron emission ($< -1.5$).

Figure~\ref{fig:phase_space} shows a clear dichotomy between shock-powered synchrotron transients with mildly relativistic velocities $0.2 \lesssim \beta_{\rm sh} \lesssim 1$ and those with non-relativistic velocities $\beta_{\rm sh} \ll 1$. In the former case, peak emission is dominated by thermal electrons and a steep optically-thin spectrum can be attained, whereas the latter are governed entirely by the non-thermal power-law electron distribution. This dichotomy  almost exclusively depends on shock velocity with only very weak dependence on density. This is because the thermal optical-depth parameter $\tau_\Theta$ scales strongly with velocity (eq.~\ref{eq:tau_Theta}). 
Specifically, in order for thermal electrons to contribute to the optically-thin emission, the frequency at which emission transitions from thermal to non-thermal electrons must fall above the self-absorption frequency, i.e.,  $\nu_{\rm a,th} < \nu_j$ must be satisfied.   For our fiducial $\delta = 0.01$ this implies (eqs.~\ref{eq:xj},\ref{eq:xa_th_slowcooling},\ref{eq:xa_th_fastcooling}) $\tau_\Theta < 1.7 \times 10^9$ (or $\Theta \tau_\Theta/\gamma_{\rm cool} < 2 \times 10^{10}$ in the fast-cooling regime), and therefore that
\begin{equation}
\label{eq:th_electron_condition}
    \beta_{\rm sh} 
    \underset{\delta=0.01}{\gtrsim}
    \min
    \begin{cases}
    0.22 \, \epsilon_T^{-1/2} \epsilon_{B,-1}^{-1/20} n_5^{1/20} t_{100}^{-1/10} 
    \\
    0.15 \, \epsilon_T^{-3/7} \epsilon_{B,-1}^{-3/28} n_5^{-1/28} 
    \end{cases}
\end{equation}
is required for thermal electrons to dominate the SED peak. 
The top case corresponds to the slow-cooling regime while the bottom case applies in the fast-cooling regime ($\nu_{\rm a,th} < \nu_{\rm cool,th}$).
Smaller (larger) values of $\delta$ would imply a lower (higher) threshold velocity.
Specifically, using the rough scalings $x_j \propto \delta^{-0.25}$ and $x_a \propto \tau_\Theta^{0.2}$ (see text below eqs.~\ref{eq:xj},\ref{eq:xa_th_slowcooling}) we find that the critical shock velocity (eq.~\ref{eq:th_electron_condition}) scales as $\delta^{0.125}$ in the slow cooling regime, and $\delta^{0.089}$ in the fast-cooling case.

Condition~(\ref{eq:th_electron_condition}) also reflects the thermal electron temperature $\Theta$, shown with the right vertical axis in Fig~\ref{fig:phase_space}. When $\Theta \gtrsim 1$ thermal electrons are relativistic and produce copious synchrotron emission, whereas the majority of thermal electrons are non-relativistic if $\Theta \ll 1$ and only a small fraction are capable of contributing to emission at frequencies $\nu \gg \nu_\Theta$ of relevance.
The threshold velocity (eq.~\ref{eq:th_electron_condition}) therefore depends on the electron thermalization efficiency $\epsilon_T$. Lower efficiencies (smaller $\epsilon_T$) would increase the threshold shock velocity and push the region where thermal electrons dominate peak emission 
to higher shock velocities.

The strong dependence on shock velocity evident in Fig.~\ref{fig:phase_space} helps explain why non-relativistic shocks in radio SNe are well modelled by a power-law electron distribution and do not show any clear evidence for thermal electrons, whereas the emerging class of AT2018cow-like events that have mildly-relativistic inferred velocities $\beta_{\rm sh} \gtrsim 0.2$ exhibit steep spectra consistent with a contribution from thermal electrons \citep{Ho+21b}.
Red points in Fig.~\ref{fig:phase_space} show shock properties inferred by \cite{Ho+21b} for AT2018cow at an epoch of 10\,d \citep{Margutti+19,Ho+19}, AT2020xnd at 40\,d \citep{Ho+21b}, CSS161010 at 99\,d \citep{Coppejans+20}, and AT2018lug at 81\,d (the `Koala'; \citealt{Ho+20}). In comparison, typical radio SNe have velocities of order $\beta_{\rm sh} \sim 0.03$ and densities $n \sim 10^5-10^6 \, {\rm cm}^{-3}$ at timescales of $\sim 100\,{\rm d}$ post-explosion \citep{Weiler+02}.

In addition to providing a natural explanation for why steep thermal-electron spectra would be seen in AT2018cow-like events but not in standard radio SNe, Fig.~\ref{fig:phase_space} may also help explain the unusually bright and prolonged millimeter emission observed in AT2018cow and AT2020xnd. At a fixed ambient density---shocks with higher velocities produce more luminous emission that peaks at higher frequencies (especially above $\beta_{\rm sh} \gtrsim 0.1$ where the blue contours kink to the left). In particular, there is a large swath of parameter-space where emission peaks in the millimeter band. This is especially pronounced considering potential selection biases towards detecting the most luminous events.

\section{Temporal Evolution}
\label{sec:time_evolution}

In Fig.~\ref{fig:phase_space} we presented the phase-space of synchrotron-powered transients as a function of shock velocity and ambient density, at a fixed epoch $t$. Here we briefly discuss the temporal evolution of transients within this phase-space, as time (and potentially upstream density, shock velocity) progresses. There is a rich phenomenology of possible light-curves depending on the time evolution of various quantities of interest. 
Here we focus on the specific case where the upstream follows a wind density profile, $n \propto r^{-2}$.

\begin{figure}
    \includegraphics[width=0.5\textwidth]{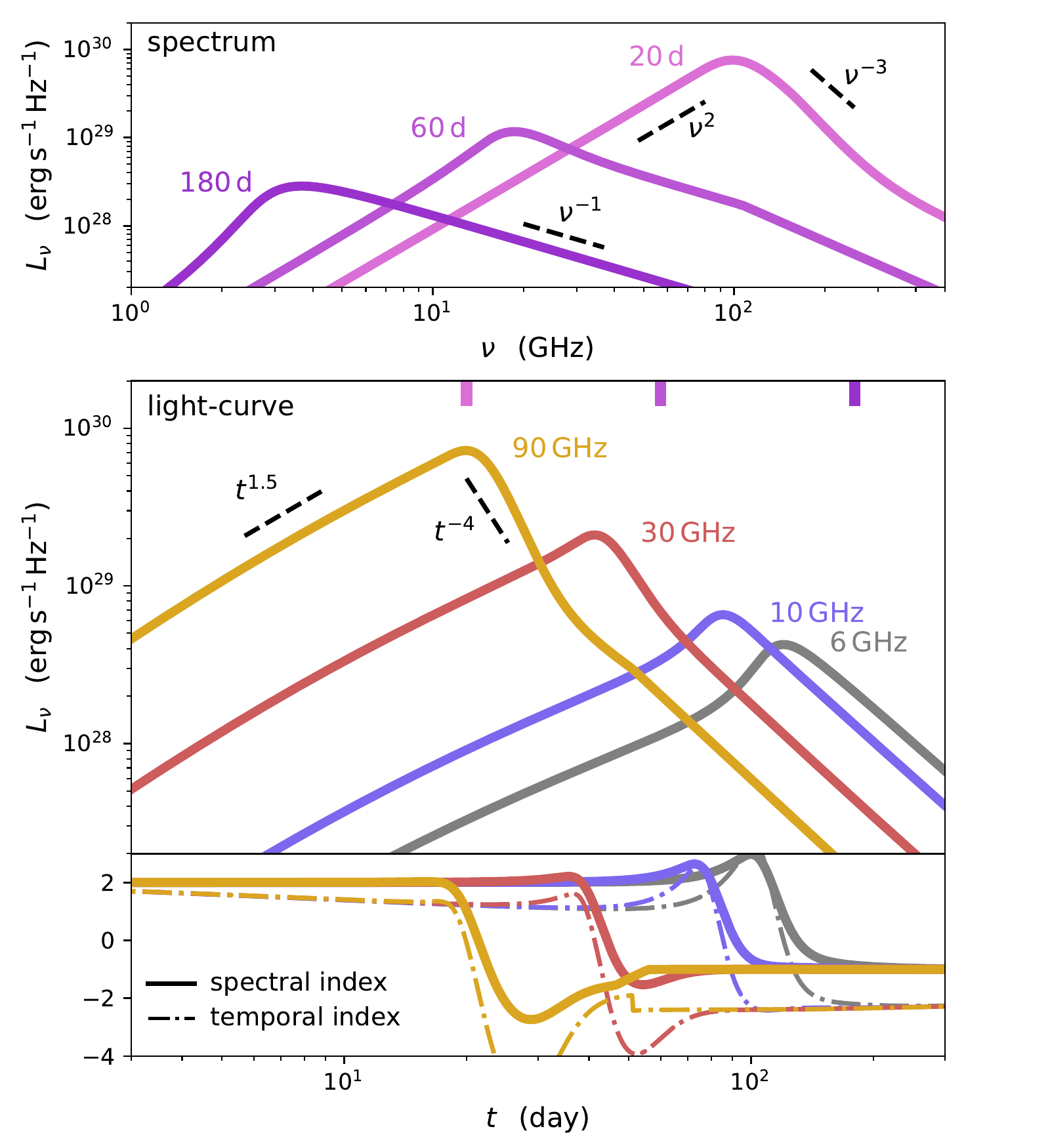}
    \caption{{\it Middle Panel:} Example light-curves for a decelerating shock-wave in a wind density profile. At a fixed frequency (labeled), the light-curve peaks when the SSA frequency passes through the band. The high-frequency peak is dominated by thermal electrons. This implies a steep post-peak decline, qualitatively consistent with observed AT2018cow-like events (eq.~\ref{eq:lightcurve_decline}). 
    At later times the light-curve samples power-law electrons and the decline-rate softens.
    {\it Top:} SED snapshots of the same model at different epochs. At early times the SED peaks at high-frequencies showing tell-tale signs of thermal electrons---a steep optically-thin spectrum (eq.~\ref{eq:th_spectral_index}) and a $\propto \nu^2$ self-absorbed slope. Shock deceleration causes the observed contribution of thermal electrons to drop with time. By 180\,d the SED lacks clear signatures of thermal electrons and is instead governed by power-law electrons.
    {\it Bottom:} The spectral (solid) and temporal (dot-dashed) indices at different frequencies (following color-scheme of middle panel) as a function of time. At high-frequencies, both spectral and temporal indices attain steep (negative) values shortly after light-curve peak. This is a unique feature of the thermal-electron model.
    }
    \label{fig:light_curves}
\end{figure}

Figure~\ref{fig:light_curves} shows example light-curves resulting from our model. These are calculated assuming that a blast-wave of initial velocity $\beta_{\rm sh} = 0.4$ and total energy $10^{50}\,{\rm erg}$ is driven into an ambient wind whose density is $n = 10^5\,{\rm cm}^{-3} \left(r/10^{16}\,{\rm cm}\right)^{-2}$.\footnote{
This corresponds to a mass-loss rate of $\dot{M} \simeq 2.1 \times 10^{-4} \,M_\odot\, {\rm yr}^{-1}\, \left(v_{\rm w}/1000\,{\rm km\,s}^{-1}\right)$, where $v_{\rm w}$ is the wind velocity.
}
The shock dynamics are integrated assuming a spherical thin-shell model that is accurate in both the relativistic and non-relativistic regimes (\citealt{Huang+99,Peer12}; see \citealt{Schroeder+20} for details on this implementation).
This results in a gradually decelerating shock and a time-dependent shock velocity $c \beta_{\rm sh}(t)$, radius $R(t)$, and upstream density $n[R(t)]$.
We calculate the light-curves from eq.~(\ref{eq:Lnu}) using these time-dependent quantities and adopting fiducial $\epsilon_T=1$, $\epsilon_B=0.1$, $\delta = 0.01$, and $p=3$.

The middle panel of Fig.~\ref{fig:light_curves} shows resulting light-curves at different frequencies (labeled). At a given frequency, the light-curve peaks when the SSA frequency passes through the band.
Shortly after peak, the high-frequency light-curves exhibit a sharp drop. This is directly related to the steep thermal spectrum at frequencies $\nu > \nu_{\rm a,th}$ (eq.~\ref{eq:th_spectral_index}) and is a unique property of the thermal electron model. The top panel of Fig.~\ref{fig:light_curves} shows snapshots of the spectrum at different epochs and illustrates the steep optically-thin SED that is obtained at early times.
The correlation between the unusually steep spectrum and steep light-curve decline rate is further illustrated by the bottom panel in this figure, which shows the spectral index ($d\ln L_\nu / d\ln \nu$; solid curves) and temporal index ($d\ln L_\nu / d\ln t$; dot-dashed curves) at different frequencies (different colors). Both the temporal and spectral indices obtain steep (negative) values shortly after peak at $90\,{\rm GHz}$.

Figure~\ref{fig:light_curves} illustrates another important feature:  if there is enough mass in the surrounding CSM, shock deceleration will eventually cause initially mildly-relativistic shocks that satisfy eq.~(\ref{eq:th_electron_condition}) to violate this condition at late times. This implies that the relative contribution of thermal electrons to the observed emission will decay as a function of time, and that at late enough epochs the light-curves and spectra will revert to the standard power-law electron distribution picture.
This can be seen from the SED snapshots in the top panel of Fig.~\ref{fig:light_curves}. At early times, the spectrum exhibits the tell-tale $\nu^2$ self-absorbed rise and steep optically-thin decline that are characteristic of thermal electrons (see left panel, Fig.~\ref{fig:SEDs}). At later epochs the optically-thin (optically-thick) slope flattens (steepens) and is eventually governed entirely by non-thermal electrons.
This is also imprinted in the late-time low-frequency light-curves, that no longer show the steep post-peak decline apparent at higher-frequencies.
We note that this agrees with modeling of AT2018cow and AT2020xnd, which suggested that the late-time data was well-fit within the standard power-law synchrotron framework \citep{Margutti+19,Ho+21b}.
The spectrum in this power-law dominated regime follows the right panel in Fig.~\ref{fig:SEDs}, and the temporal evolution can be derived using eqs.~(\ref{eq:Lnu},\ref{eq:xa_pl}). This reverts to the results of \cite{Chevalier98} in the slow-cooling regime, and to the results presented in Appendix~C of \cite{Ho+21b} for the fast-cooling regime.

We can understand the results presented in Fig.~\ref{fig:light_curves} more quantitatively by estimating the light-curve scalings in the case where thermal electrons dominate the emission (as particularly relevant at high-frequencies and early epochs).
We pursue this by denoting the temporal-scaling of the shock radius as $R \propto t^m$. 
In general (and in our numerical model) the radius does not follow a power-law evolution and the exponent $m$ should instead be interpreted as the instantaneous expansion rate $d\ln R/d\ln t$. In typical cases we expect $m=1$ at early epochs before significant shock deceleration, and lower values of $m$ at later times (the Sedov-Taylor solution for a wind medium sets a lower limit of $m \geq 2/3$).
This scaling implies that $\beta_{\rm sh} \propto t^{m-1}$ and $n \propto t^{-2m}$.
If the electron temperature and magnetic field are determined by eqs.~(\ref{eq:Theta},\ref{eq:Theta_0},\ref{eq:B}) then we additionally have $\Theta \propto t^{2(m-1)}$, and $B \propto t^{-1}$.

The SED peak is set by SSA. If this is governed by thermal electrons then the SSA frequency is given by eqs.~(\ref{eq:nu_Theta},\ref{eq:xa_th_slowcooling},\ref{eq:xa_th_fastcooling}). Here we adopt the simpler approximation $x_{\rm a,th} \propto \tau_\Theta^{0.2}$ (or $x_{\rm a,th} \propto (\tau_\Theta \gamma_{\rm cool}/\Theta)^{0.2}$ for fast-cooling electrons) that is better-suited for deriving analytic scaling relations.
Using these approximations along with eqs.~(\ref{eq:nu_Theta},\ref{eq:tau_Theta},\ref{eq:gamma_cool}), we find that the thermal SSA frequency scales as
\begin{equation}
\label{eq:nua_th_scaling}
    \nu_{\rm a,th} \propto
    \begin{cases}
    t^{1.8m-2.8} &, 
    \nu_{\rm a,th} < \nu_{\rm cool}
    \\
    t^{1.4m-2.2} &, 
    \nu_{\rm a,th} > \nu_{\rm cool}
    \end{cases}
\end{equation}
for a wind density profile and $\Theta \gtrsim 1$.
The top (slow-cooling) case is bound between $t^{-1}$ to $t^{-1.6}$ for physical values of $2/3 \leq m \leq 1$.
The peak luminosity of SSA thermal electrons is $L_{\rm a,th} \propto R^2 \nu_{\rm a,th}^2 \Theta$ (eq.~\ref{eq:Lnu}), which therefore scales as 
\begin{equation}
\label{eq:La_th_scaling}
    L_{\rm a,th} \propto
    \begin{cases}
    t^{-7.6 (1-m)} &, \nu_{\rm a,th} < \nu_{\rm cool}
    \\
    t^{6.8m-6.4} &, \nu_{\rm a,th} > \nu_{\rm cool}
    \end{cases}
    .
\end{equation}
In the slow-cooling case this is bound between $L_{\rm a,th} \sim const$ and $L_{\rm a,th} \propto t^{-2.53}$.
The above equations imply that---for a wind density medium---the peak (SSA) frequency of thermal electrons drops moderately as a function of time, while the peak flux is sensitive to the shock deceleration parameter $m$.

If emission is dominated by thermal electrons, then $\nu < \nu_{\rm a,th}$ prior to the light-curve peak and the luminosity is $L_\nu \approx \left(\nu/\nu_{\rm a,th}\right)^2 L_{\rm a,th}$. 
From eqs.~(\ref{eq:nua_th_scaling},\ref{eq:La_th_scaling}) this implies 
\begin{equation}
    L_\nu \left(t<{\rm peak}\right)
    \propto t^{4m-2}
\end{equation}
in both the slow- and fast-cooling regimes, and that the light-curve rises to peak as $\sim t^{2/3} - t^2$. This is consistent with our numerical results shown in Fig.~\ref{fig:light_curves}. 

Following the light-curve peak, unusually steep decays were observed at high frequencies for AT2018cow and AT2020xnd, qualitatively consistent with the thermal electron model.
For example, in Fig.~\ref{fig:light_curves} we show the $\sim t^{-4}$ scaling inferred for AT2020xnd to guide the eye \citep{Ho+21b}.
A crude analytic estimate of the temporal slope in this regime can be derived using eqs.~(\ref{eq:th_spectral_index},\ref{eq:nua_th_scaling},\ref{eq:La_th_scaling}). Shortly after peak the luminosity is roughly $L_\nu \sim \left(\nu/\nu_{\rm a,th}\right)^{\alpha_{\rm th}(\nu_{\rm a,th})} L_{\rm a,th}$ and therefore
\begin{equation}
\label{eq:lightcurve_decline}
    L_\nu \left(t \gtrsim {\rm peak}\right)
    \propto
    \begin{cases}
    t^{-7.6 (1-m) - (1.8m-2.8)\alpha_{\rm th}} &, \nu_{\rm a,th} < \nu_{\rm cool}
    \\
    t^{6.8m-6.4 - (1.4m-2.2)\alpha_{\rm th}} &, \nu_{\rm a,th} > \nu_{\rm cool}
    \end{cases}
    .
\end{equation}
For example, if the spectral index is $\alpha_{\rm th} = -2$, we obtain that $L_\nu \propto t^{11.2m-13.2}$ in the slow-cooling regime. Even for very mild deceleration this implies a very steeply declining light-curve (e.g. $L_\nu \propto t^{-3}$ for $m \approx 0.9$). Steeper spectral indices and/or stronger deceleration yield light-curves that decay more abruptly.

\section{Discussion}
\label{sec:discussion}

In this work we studied the implications of a thermal electron population on sub-relativistic shock-powered synchrotron transients. 
The existence of a thermal electron population is a natural expectation in shock scenarios, yet has garnered little attention in the context of synchrotron transients. 
We find that neglecting thermal electrons is reasonable only for non-relativistic shocks where $\beta_{\rm sh} \ll 1$. Much of the canonical synchrotron transient literature was derived for radio SNe where this is applicable, however the situation is markedly different for mildly-relativistic shocks (Fig.~\ref{fig:phase_space}). 
If dominant, thermal electrons can be discerned by their tell-tale steep optically-thin spectrum (eq.~\ref{eq:th_spectral_index}) and a comparatively shallow $\nu^2$ self-absorbed spectrum (Fig.~\ref{fig:SEDs}). Another general prediction of the thermal electron model is a steep decay of the light-curve shortly after peak, and a correlation between the spectral and temporal indices (Fig.~\ref{fig:light_curves}; eq.~\ref{eq:lightcurve_decline}). In typical settings, these effects should be most prominent at early times and at high-frequencies.

The physical processes that determine the post-shock electron temperature are a matter of ongoing investigation, but it is generally recognized that plasma instabilities must mediate electron-ion energy exchange.\footnote{
The timescale for electron-ion equilibration through Coulomb collisions is typically too slow, $t_{\rm ei} \sim 200\,{\rm yr}\, n_5^{-1} \Theta^b (\ln \Lambda / 30)^{-1}$ where $\ln \Lambda$ is the Coulomb logarithm, $b=3/2$ for $\Theta \ll 1$ \citep{Spitzer56} and $b=1$ for $1 \ll \Theta \ll m_p/m_e$ \citep{Stepney83}.
} 
PIC simulations of both relativistic and sub-relativistic electron-ion shocks generically show a quasi-thermal downstream electron population that 
shares an order-unity fraction of the downstream energy ($\epsilon_T \sim 1$) and that
exceeds the energy in
the diffusive-shock-accelerated power-law tail ($\delta \ll 1$; \citealt{Sironi&Spitkovsky11,Park+15,Crumley+19,Tran&Sironi20}).
In our present work we have assumed that this quasi-thermal population can be modelled by a relativistic Maxwellian (eq.~\ref{eq:dn_dgamma_th}), i.e. that it is ``perfectly'' thermal. We expect that modest deviations from a pure Maxwell-J\"{u}ttner distribution would not affect our main conclusions, but may quantitatively change various estimates. In particular, our results are sensitive to the high-energy tail of the thermal distribution, which contributes most to emission at frequencies $\gg \nu_\Theta$ of interest.
For example, if we generalize eq.~(\ref{eq:dn_dgamma_th}) to $(dn/d\gamma)_{\rm th} \propto e^{-(\gamma/\Theta)^n}$ where $n=1$ for a standard Maxwellian, then the high frequency thermal synchrotron spectrum would scale as 
$L_\nu \propto e^{ - A_n x^{{n}/{(n+2)}} }$ with $A_n = [ 1 + ({n}/{2})^{{2}/{n}} ] (2/n)^{{n}/{(n+2)}}$ (compare with eq.~\ref{eq:I_of_x}). 
This may affect quantitative values of $\nu_j$, $\nu_\alpha$, $\nu_{\rm a,th}$, and $\nu_{\rm cool,th}$\footnote{
The cooling frequency of the thermal population would in this case be $x_{\rm cool,th} \sim (n/2) (\gamma_{\rm cool}/\Theta)^{n+2}$, affecting eqs.~(\ref{eq:cooling_correction_th},\ref{eq:nu_cool_th}).}
(eqs.~\ref{eq:nu_cool_th}--\ref{eq:xa_th_fastcooling})
but should not change our overall findings.

Our study was motivated by steep spectra and light-curves observed in several AT2018cow-like events, and by the work of \cite{Ho+21b} that first suggested a thermal-electron interpretation of this data.
Here we addressed several questions that arise from such an interpretation. We showed that thermal electrons are naturally expected to govern peak emission for mildly-relativistic shocks with $\beta_{\rm sh} \gtrsim 0.2$ (Fig.~\ref{fig:phase_space} and eq.~\ref{eq:th_electron_condition}). Conversely, thermal electrons would be subdominant for non-relativistic shocks. This explains the dichotomy between AT2018cow-like events and typical radio SNe (Fig.~\ref{fig:phase_space}).
Furthermore, we showed that the synchrotron optical depth 
at frequency $\sim \nu_\Theta$ at which most thermal electrons emit is $\gg 1$ for sub-relativistic shocks (eq.~\ref{eq:tau_Theta}). This implies that a ``bare'' (unabsorbed) Maxwellian should not be observed for such shocks, and helps explain why the SED peak in AT2018cow-like events is inferred to be $\sim \mathcal{O}(100)$ times above $\nu_\Theta$ \citep{Ho+21b}.
The thermal optical depth $\langle \alpha_{\nu,{\rm th}} \rangle R \propto \tau_\Theta e^{-1.8899 x^{1/3}}$ is exponentially sensitive to frequency, which acts to regulate the self-absorption frequency to $x_{\rm a, th} \sim \mathcal{O}(100)$ over many orders of magnitude in $\tau_\Theta$ (eq.~\ref{eq:xa_th_slowcooling}).

In addition to shock velocity, a second important parameter that determines the contribution of thermal electrons to observed emission is the ambient density. At a fixed shock velocity, the density sets the downstream magnetic field (eq.~\ref{eq:B}) and therefore governs the frequency 
$\nu_j$ below which thermal electrons dominate observed emission (Fig.~\ref{fig:SEDs}).
Using the rough approximation $x_j \approx 150 \delta^{-0.25}$ (see text below eq.~\ref{eq:xj}) and eq.~(\ref{eq:nu_Theta}), this critical frequency is
\begin{equation}
\label{eq:nu_j}
    \nu_j 
    \approx
    3 \,{\rm GHz}\, \epsilon_T^2 \epsilon_{B,-1}^{1/2} n_5^{1/2} \beta_{-1}^5
    \left({\delta}/{0.01}\right)^{-0.25}
    .
\end{equation}
At frequencies $\nu > \nu_j$ emission is dominated by non-thermal electrons and the presence of thermal electrons would be undetectable. 

Eq.~(\ref{eq:nu_j}) shows that $\nu_j$ falls in the GHz--mm band for high-density high-velocity shocks relevant to AT2018cow-like events.
Although this class of events has been our primary focus in this work, our results would apply to sub-relativistic shocks in any other astrophysical setting as well.
For example, BNS mergers eject $\sim 10^{-2}M_\odot$ of material at velocities $\gtrsim 0.1 c$, and it has been suggested that the forward shock between this ejecta and the ambient interstellar medium (ISM) would produce detectable synchrotron radio emission \citep[e.g.][]{Nakar&Piran11,Margalit&Piran15,Margalit&Piran20,Hajela+21}. This has typically been studied using standard power-law electron models, however---as we have shown---thermal electrons may contribute appreciably for mildly-relativistic shocks. 
This contribution is limited to low frequencies 
$\nu < \nu_j \approx 300\,{\rm MHz}\,\epsilon_{B,-1}^{1/2} n_0^{1/2} (\beta_{\rm sh}/0.2)^5$ (eq.~\ref{eq:nu_j}) that may hinder detectability prospects, especially considering that the ambient ISM density is likely to be significantly lower than the optimistic value assumed above \citep[e.g.][]{Fong+15,Hajela+19}. Nevertheless, our estimate motivates late-time follow-up of BNS mergers at particularly low frequencies in order to test the thermal-electron hypothesis.
Furthermore, if the merger manages to produce a ``long-lived'' magnetar remnant then energy injection may accelerate the BNS-merger ejecta to trans-relativistic velocities \citep{Metzger&Bower14,Horesh+16,Fong+16,Margalit&Metzger19,Schroeder+20} implying much higher frequencies up to which thermal electrons may contribute noticeably (eq.~\ref{eq:nu_j}).

GRB afterglows could show signs of thermal electrons,
as first discussed by \cite{Giannios&Spitkovsky09}. GRB outflows are initially ultra-relativistic and collimated, however shock deceleration leads the outflow to a quasi-spherical sub-relativistic state at sufficiently late times. At such epochs, our current formalism applies (see \citealt{Ressler&Laskar17} for treatment of the ultra-relativistic regime) and we estimate
$\nu_j \approx 2.7\,{\rm GHz}\,\epsilon_{B,-1}^{1/2} E_{50} n_0^{-1/2} (t/{\rm yr})^{-3}$
for $\delta=0.01$, and for a GRB of total energy $E=E_{50}10^{50}\,{\rm erg}$ that is deep within the Sedov-Taylor regime. Observations at low frequencies $\nu < \nu_j$ would be required to potentially distinguish the thermal-electron model from a purely power-law electron distribution. Since the critical frequency $\nu_j$ drops rapidly with time, the most opportune window would be to observe shortly after the shock enters the mildly relativistic regime (the earliest epoch at which our sub-relativistic results apply).
Thermal electrons may also be relevant to other astrophysical settings in which mildly-relativistic shocks are present. This may apply to jetted tidal-disruption events such as the prototypical {\it Swift} J1644 \citep{Bloom+11,Burrows+11,Zauderer+11,Eftekhari+18}, to low-luminosity GRBs \citep[e.g.][]{Kulkarni+98,Tan+01,BarniolDuran+15}, and perhaps to more exotic scenarios such as outflows from accretion-induced collapse \citep[e.g.][]{Dessart+06,Darbha+10}.

The most important quantity that governs thermal electron emission is the post-shock electron temperature $\Theta$. In our current model, this is set uniquely by the shock velocity (eqs.~\ref{eq:Theta},\ref{eq:Theta_0}), however additional processes may impact this result. For example, inverse-Compton scattering off external (or self-produced) photons could potentially cool post-shock electrons \citep[e.g.][]{Katz+11,Margalit+21}, though this would have to act over extremely short timescales to compete with kinetic instabilities and regulate $\Theta$.\footnote{Separately from this, inverse-Compton scattering may affect the cooling break (eq.~\ref{eq:gamma_cool}) if the radiation energy density is $\gtrsim B^2/8\pi$.}

We have worked here under the simplest hypothesis that the thermal and non-thermal electrons can be adequately described by fixed values of the parameters $\epsilon_T$, $\delta$, and $p$. The microphysics that sets these parameters may in reality be more complex. For example, the strength of the power-law component $\delta$ (which is $\propto \epsilon_e$ in standard notation\footnote{
$\delta = \epsilon_e/\epsilon_T$ if $\epsilon_e$ 
is defined using energy of 
power-law electrons with $\gamma \geq \gamma_m$ (eq.~\ref{eq:gamma_m}). In practice, $\epsilon_e$ is often measured starting at some higher Lorentz factor $\gamma_t$ (e.g. at the transition between thermal and non-thermal populations, as in \citealt{Giannios&Spitkovsky09}). 
In this case $\delta$ would be a factor $(\gamma_t/\gamma_m)^{p-2} \epsilon_T^{-1}$ larger than $\epsilon_e$.
}) may itself be affected by the shock velocity. 
PIC simulations have found that $\epsilon_e \sim 0.1$ for ultra-relativistic electron-ion shocks \citep{Sironi&Spitkovsky11} while $\epsilon_e \sim 10^{-4}$ for non-relativistic shocks \citep{Park+15} and a trend of increasing $\epsilon_e$ with shock velocity has been suggested \citep[e.g.][]{Crumley+19}.
If this trend is correct, then the parameter space in Fig.~\ref{fig:phase_space} where thermal electrons contribute appreciably would expand and encompass even lower-velocity shocks. This would improve prospects for detecting emission from such thermal electrons, but may already be at odds with observations of events straddling the two regions. 

The shock Mach-number and magnetic field orientation can further affect diffusive-shock acceleration and impact $\delta$ and/or $p$. For example, shocks where the magnetic field is perpendicular to the shock velocity are thought to be less efficient at accelerating non-thermal particles \citep[e.g.][]{Sironi&Spitkovsky09} and may produce more prominent thermal downstream distributions, i.e. $\delta \ll 1$ (although see \citealt{Xu+20,Kumar&Reville21}). This would again expand the range of parameters where thermal electrons must be considered.

The above uncertainties in microphysics (that are in any case not considered in typical modelling of synchrotron transients) should not be considered detrimental to the thermal + non-thermal model. In fact, we view these as an important opportunity---our model provides a direct means of measuring the acceleration efficiency $\delta$ and therefore constraining microphysical processes using observations. Such direct measurement is possible if one observes the transition frequency $\nu_j$ (eq.~\ref{eq:nu_j}) at which emission changes from thermal to power-law dominated (or similarly by detecting $\nu_\alpha$ in the SSA regime).
\cite{Ho+21b} have already used this method to infer that $\delta < 0.2$ for AT2020xnd.

Finally, we note that the thermal + non-thermal synchrotron model presented here includes no additional physical parameters compared to standard non-thermal synchrotron models that are typically used to model observations. At a given epoch, the frequency-dependent spectral luminosity is fully specified by three physical parameters, $R$, $\beta_{\rm sh}$, and $n$, and three microphysical parameters, $\epsilon_B$, $\delta$, and $p$ (insofar as $\epsilon_T \sim 1$, which is well-motivated in the simplest version of this model). The same parameters are also required in standard non-thermal synchrotron modelling (with $\epsilon_e$ replacing $\delta$). The fact that this model is capable of fitting more complex spectra \citep[see][]{Ho+21b} with no additional degrees of freedom is another strength of this scenario.
In the Appendix we provide analytic expressions (that are applicable in a subset of the parameter space) and a link to the code used in our analysis (applicable for any sub-relativistic shocks). These may be convenient for future studies, and in particular for fitting observed data to the model presented in this work.

\acknowledgements
We thank Anna Ho for helpful comments and discussions on AT2020xnd, AT2018cow, and radio SNe, and Lorenzo Sironi and Anatoly Spitkovsky for useful discussions about collisionless shocks.  BM is supported by NASA through the NASA Hubble Fellowship grant \#HST-HF2-51412.001-A awarded by the Space Telescope Science Institute, which is operated by the Association of Universities for Research in Astronomy, Inc., for NASA, under contract NAS5-26555.   EQ was supported in part by a Simons Investigator grant from the Simons Foundation.

\appendix
\section{Fitting the Model to Observed Data}

The code used in this analysis is publicly available at 
\href{https://github.com/bmargalit/thermal-synchrotron}{https://github.com/bmargalit/thermal-synchrotron}. 
It is designed to be imported as a python module and can easily be used to calculate the specific luminosity (eq.~\ref{eq:Lnu}) for any set of parameters. This may be particularly useful for fitting observed data to the thermal + non-thermal model described in this work.
In the following, we also provide analytic expressions that can be used to fit observed light-curves and SEDs in the reduced case where $\Theta \gtrsim 1$. For concreteness, we also choose $p=3$ in the expressions below (the linked code can be used for arbitrary values of $\Theta$ and $p$). In these limits, the specific luminosity (eq.~\ref{eq:Lnu}) can be expressed as
\begin{equation}
\label{eq:Appendix_Lnu_fit}
    L_\nu(\nu) 
    \underset{\substack{p=3 \\ \Theta \gtrsim 1}}{=}
    L_\Theta
    \left[ x I^\prime(x) \min\left(1, \frac{z_{\rm cool}}{(0.5x)^{1/3}}\right) + 9.674 \times \delta x^{-1} \min\left(1, \frac{z_{\rm cool}}{x^{1/2}}\right)
    \right]
    \frac{1-e^{-\tau(x)}}{\tau(x)}
\end{equation}
where $x \equiv \nu / \nu_\Theta$ is the normalized frequency (eq.~\ref{eq:nu_Theta}), $L_\Theta$ is a normalization constant, $I^\prime(x)$ is given by eq.~(\ref{eq:I_of_x}; see also \citealt{Mahadevan+96}),
\begin{equation}
\label{eq:Appendix_tau}
    \tau(x) = \tau_\Theta 
    \left[
    x^{-1} I^\prime(x) \min\left(1, \frac{z_{\rm cool}}{(0.5x)^{1/3}}\right) + 47.37 \times \delta x^{-7/2} \min\left(1, \frac{z_{\rm cool}}{x^{1/2}}\right)
    \right]
\end{equation}
is the frequency-dependent optical depth $\propto \tau_\Theta$ (eq.~\ref{eq:tau_Theta}), and $z_{\rm cool} \equiv \gamma_{\rm cool}/\Theta$ is related to the synchrotron cooling break (eq.~\ref{eq:gamma_cool}).
Eq.~(\ref{eq:Appendix_Lnu_fit}) can be fit to observed spectra as a function of the five parameters: $L_\Theta$, $\tau_\Theta$, $\nu_\Theta$, $\delta$, and $z_{\rm cool}$. These may subsequently be related to the shock velocity $\beta_{\rm sh}$, upstream density $n$, and radius $R$ using eqs.~(\ref{eq:nu_Theta},\ref{eq:tau_Theta},\ref{eq:Lnu}),
\begin{equation}
    \beta_{\rm sh} \approx 0.276 \, 
    \left(\frac{L_\Theta}{10^{35}\,{\rm erg\,s}^{-1}\,{\rm Hz}^{-1}}\right)^{1/34} \left(\frac{\tau_\Theta}{10^8}\right)^{-3/34} \epsilon_{B,-1}^{-1/17} \epsilon_T^{-15/34}
    ,
\end{equation}
\begin{equation}
    n \approx 1.3 \times 10^5 \,{\rm cm}^{-3}\,
    \left(\frac{\nu_\Theta}{1\,{\rm GHz}}\right)^{2} \left(\frac{L_\Theta}{10^{35}\,{\rm erg\,s}^{-1}\,{\rm Hz}^{-1}}\right)^{-5/17} \left(\frac{\tau_\Theta}{10^8}\right)^{15/17} \epsilon_{B,-1}^{-7/17} \epsilon_T^{7/17}
    ,
\end{equation}
\begin{equation}
    R \approx 4.5 \times 10^{16} \,{\rm cm}\,
    \left(\frac{\nu_\Theta}{1\,{\rm GHz}}\right)^{-1} \left(\frac{L_\Theta}{10^{35}\,{\rm erg\,s}^{-1}\,{\rm Hz}^{-1}}\right)^{8/17} \left(\frac{\tau_\Theta}{10^8}\right)^{-7/17} \epsilon_{B,-1}^{1/17} \epsilon_T^{-1/17}
    .
\end{equation}
Note that the three parameters $L_\Theta$, $\tau_\Theta$, and $\nu_\Theta$ are sufficient to fully specify the physical shock parameters. If the power-law electron distribution does not impact observed frequencies ($\nu < \nu_j$) then $\delta$ can effectively be set to $\delta = 0$. Similarly, if fast-cooling is irrelevant at frequencies of interest ($\nu < \nu_{\rm cool}$) then one can effectively set $z_{\rm cool} \to \infty$. In these cases eqs.~(\ref{eq:Appendix_Lnu_fit},\ref{eq:Appendix_tau}) reduce to a form similar to that used by \cite{Ho+21b} in fitting the SEDs of AT2020xnd, AT2018cow, and CSS161010.
Alternatively, if the cooling-break parameter $z_{\rm cool}$ can be constrained by the SED fit then the expressions above can be combined with eq.~(\ref{eq:gamma_cool}) to directly measure the microphysical parameter $\epsilon_B$,
\begin{equation}
    \epsilon_B \approx 8.2 \times 10^{-3} \,
    \left(\frac{z_{\rm cool}}{10}\right)^{-1} \left(\frac{\nu_\Theta}{1\,{\rm GHz}}\right)^{-1} \left(\frac{L_\Theta}{10^{35}\,{\rm erg\,s}^{-1}\,{\rm Hz}^{-1}}\right)^{-9/34} \left(\frac{\tau_\Theta}{10^8}\right)^{-7/34} \epsilon_T^{-1/34}
    .
\end{equation}
Similarly, if power-law electrons contribute to observed emission then the microphysical parameter $\delta$ may be directly constrained by fitting eq.~(\ref{eq:Appendix_Lnu_fit}) to the data.


\bibliography{bib}{}

\begin{thebibliography}{}
\expandafter\ifx\csname natexlab\endcsname\relax\def\natexlab#1{#1}\fi
\providecommand{\url}[1]{\href{#1}{#1}}
\providecommand{\dodoi}[1]{doi:~\href{http://doi.org/#1}{\nolinkurl{#1}}}
\providecommand{\doeprint}[1]{\href{http://ascl.net/#1}{\nolinkurl{http://ascl.net/#1}}}
\providecommand{\doarXiv}[1]{\href{https://arxiv.org/abs/#1}{\nolinkurl{https://arxiv.org/abs/#1}}}

\bibitem[{{Barniol Duran} {et~al.}(2015){Barniol Duran}, {Nakar}, {Piran}, \&
  {Sari}}]{BarniolDuran+15}
{Barniol Duran}, R., {Nakar}, E., {Piran}, T., \& {Sari}, R. 2015, \mnras, 448,
  417, \dodoi{10.1093/mnras/stv011}

\bibitem[{{Bell}(1978)}]{Bell78}
{Bell}, A.~R. 1978, \mnras, 182, 147, \dodoi{10.1093/mnras/182.2.147}

\bibitem[{{Blandford} \& {Eichler}(1987)}]{BlandfordEichler87}
{Blandford}, R., \& {Eichler}, D. 1987, \physrep, 154, 1,
  \dodoi{10.1016/0370-1573(87)90134-7}

\bibitem[{{Blandford} \& {Ostriker}(1978)}]{Blandford&Ostriker78}
{Blandford}, R.~D., \& {Ostriker}, J.~P. 1978, \apjl, 221, L29,
  \dodoi{10.1086/182658}

\bibitem[{{Bloom} {et~al.}(2011){Bloom}, {Giannios}, {Metzger}, {Cenko},
  {Perley}, {Butler}, {Tanvir}, {Levan}, {O'Brien}, {Strubbe}, {De Colle},
  {Ramirez-Ruiz}, {Lee}, {Nayakshin}, {Quataert}, {King}, {Cucchiara},
  {Guillochon}, {Bower}, {Fruchter}, {Morgan}, \& {van der Horst}}]{Bloom+11}
{Bloom}, J.~S., {Giannios}, D., {Metzger}, B.~D., {et~al.} 2011, Science, 333,
  203, \dodoi{10.1126/science.1207150}

\bibitem[{{Bright} {et~al.}(2021){Bright}, {Margutti}, {Matthews}, {Brethauer},
  {Coppejans}, {Wieringa}, {Metzger}, {DeMarchi}, {Laskar}, {Romero},
  {Alexander}, {Horesh}, {Migliori}, {Chornock}, {Berger}, {Bietenholz},
  {Devlin}, {Dicker}, {Jacobson-Gal{\'a}n}, {Mason}, {Milisavljevic}, {Motta},
  {Mroczkowski}, {Ramirez-Ruiz}, {Rhodes}, {Sarazin}, {Sfaradi}, \&
  {Sievers}}]{Bright+21}
{Bright}, J.~S., {Margutti}, R., {Matthews}, D., {et~al.} 2021, arXiv e-prints,
  arXiv:2110.05514.
\newblock \doarXiv{2110.05514}

\bibitem[{{Burrows} {et~al.}(2011){Burrows}, {Kennea}, {Ghisellini}, {Mangano},
  {Zhang}, {Page}, {Eracleous}, {Romano}, {Sakamoto}, {Falcone}, {Osborne},
  {Campana}, {Beardmore}, {Breeveld}, {Chester}, {Corbet}, {Covino},
  {Cummings}, {D'Avanzo}, {D'Elia}, {Esposito}, {Evans}, {Fugazza}, {Gelbord},
  {Hiroi}, {Holland}, {Huang}, {Im}, {Israel}, {Jeon}, {Jeon}, {Jun}, {Kawai},
  {Kim}, {Krimm}, {Marshall}, {P. M{\'e}sz{\'a}ros}, {Negoro}, {Omodei},
  {Park}, {Perkins}, {Sugizaki}, {Sung}, {Tagliaferri}, {Troja}, {Ueda},
  {Urata}, {Usui}, {Antonelli}, {Barthelmy}, {Cusumano}, {Giommi}, {Melandri},
  {Perri}, {Racusin}, {Sbarufatti}, {Siegel}, \& {Gehrels}}]{Burrows+11}
{Burrows}, D.~N., {Kennea}, J.~A., {Ghisellini}, G., {et~al.} 2011, \nat, 476,
  421, \dodoi{10.1038/nature10374}

\bibitem[{{Chevalier}(1982)}]{Chevalier82}
{Chevalier}, R.~A. 1982, \apj, 259, 302, \dodoi{10.1086/160167}

\bibitem[{{Chevalier}(1998)}]{Chevalier98}
---. 1998, \apj, 499, 810, \dodoi{10.1086/305676}

\bibitem[{{Coppejans} {et~al.}(2020){Coppejans}, {Margutti}, {Terreran},
  {Nayana}, {Coughlin}, {Laskar}, {Alexander}, {Bietenholz}, {Caprioli},
  {Chandra}, {Drout}, {Frederiks}, {Frohmaier}, {Hurley}, {Kochanek},
  {MacLeod}, {Meisner}, {Nugent}, {Ridnaia}, {Sand}, {Svinkin}, {Ward}, {Yang},
  {Baldeschi}, {Chilingarian}, {Dong}, {Esquivia}, {Fong}, {Guidorzi},
  {Lundqvist}, {Milisavljevic}, {Paterson}, {Reichart}, {Shappee}, {Stroh},
  {Valenti}, {Zauderer}, \& {Zhang}}]{Coppejans+20}
{Coppejans}, D.~L., {Margutti}, R., {Terreran}, G., {et~al.} 2020, \apjl, 895,
  L23, \dodoi{10.3847/2041-8213/ab8cc7}

\bibitem[{{Crumley} {et~al.}(2019){Crumley}, {Caprioli}, {Markoff}, \&
  {Spitkovsky}}]{Crumley+19}
{Crumley}, P., {Caprioli}, D., {Markoff}, S., \& {Spitkovsky}, A. 2019, \mnras,
  485, 5105, \dodoi{10.1093/mnras/stz232}

\bibitem[{{Darbha} {et~al.}(2010){Darbha}, {Metzger}, {Quataert}, {Kasen},
  {Nugent}, \& {Thomas}}]{Darbha+10}
{Darbha}, S., {Metzger}, B.~D., {Quataert}, E., {et~al.} 2010, \mnras, 409,
  846, \dodoi{10.1111/j.1365-2966.2010.17353.x}

\bibitem[{{Dessart} {et~al.}(2006){Dessart}, {Burrows}, {Ott}, {Livne}, {Yoon},
  \& {Langer}}]{Dessart+06}
{Dessart}, L., {Burrows}, A., {Ott}, C.~D., {et~al.} 2006, \apj, 644, 1063,
  \dodoi{10.1086/503626}

\bibitem[{{Eftekhari} {et~al.}(2018){Eftekhari}, {Berger}, {Zauderer},
  {Margutti}, \& {Alexander}}]{Eftekhari+18}
{Eftekhari}, T., {Berger}, E., {Zauderer}, B.~A., {Margutti}, R., \&
  {Alexander}, K.~D. 2018, \apj, 854, 86, \dodoi{10.3847/1538-4357/aaa8e0}

\bibitem[{{Eichler} \& {Waxman}(2005)}]{Eichler&Waxman05}
{Eichler}, D., \& {Waxman}, E. 2005, \apj, 627, 861, \dodoi{10.1086/430596}

\bibitem[{{Fong} {et~al.}(2015){Fong}, {Berger}, {Margutti}, \&
  {Zauderer}}]{Fong+15}
{Fong}, W., {Berger}, E., {Margutti}, R., \& {Zauderer}, B.~A. 2015, \apj, 815,
  102, \dodoi{10.1088/0004-637X/815/2/102}

\bibitem[{{Fong} {et~al.}(2016){Fong}, {Metzger}, {Berger}, \&
  {{\"O}zel}}]{Fong+16}
{Fong}, W., {Metzger}, B.~D., {Berger}, E., \& {{\"O}zel}, F. 2016, \apj, 831,
  141, \dodoi{10.3847/0004-637X/831/2/141}

\bibitem[{{Gammie} \& {Popham}(1998)}]{Gammie&Popham98}
{Gammie}, C.~F., \& {Popham}, R. 1998, \apj, 498, 313, \dodoi{10.1086/305521}

\bibitem[{{Giannios} \& {Spitkovsky}(2009)}]{Giannios&Spitkovsky09}
{Giannios}, D., \& {Spitkovsky}, A. 2009, \mnras, 400, 330,
  \dodoi{10.1111/j.1365-2966.2009.15454.x}

\bibitem[{{Hajela} {et~al.}(2019){Hajela}, {Margutti}, {Alexander},
  {Kathirgamaraju}, {Baldeschi}, {Guidorzi}, {Giannios}, {Fong}, {Wu},
  {MacFadyen}, {Paggi}, {Berger}, {Blanchard}, {Chornock}, {Coppejans},
  {Cowperthwaite}, {Eftekhari}, {Gomez}, {Hosseinzadeh}, {Laskar}, {Metzger},
  {Nicholl}, {Paterson}, {Radice}, {Sironi}, {Terreran}, {Villar}, {Williams},
  {Xie}, \& {Zrake}}]{Hajela+19}
{Hajela}, A., {Margutti}, R., {Alexander}, K.~D., {et~al.} 2019, \apjl, 886,
  L17, \dodoi{10.3847/2041-8213/ab5226}

\bibitem[{{Hajela} {et~al.}(2021){Hajela}, {Margutti}, {Bright}, {Alexander},
  {Metzger}, {Nedora}, {Kathirgamaraju}, {Margalit}, {Radice}, {Berger},
  {MacFadyen}, {Giannios}, {Chornock}, {Heywood}, {Sironi}, {Gottlieb},
  {Coppejans}, {Laskar}, {Cendes}, {Barniol Duran}, {Eftekhari}, {Fong},
  {McDowell}, {Nicholl}, {Xie}, {Zrake}, {Bernuzzi}, {Broekgaarden},
  {Kilpatrick}, {Terreran}, {Villar}, {Blanchard}, {Gomez}, {Hosseinzadeh},
  {Matthews}, \& {Rastinejad}}]{Hajela+21}
{Hajela}, A., {Margutti}, R., {Bright}, J.~S., {et~al.} 2021, arXiv e-prints,
  arXiv:2104.02070.
\newblock \doarXiv{2104.02070}

\bibitem[{{Ho} {et~al.}(2019){Ho}, {Goldstein}, {Schulze}, {Khatami}, {Perley},
  {Ergon}, {Gal-Yam}, {Corsi}, {Andreoni}, {Barbarino}, {Bellm},
  {Blagorodnova}, {Bright}, {Burns}, {Cenko}, {Cunningham}, {De}, {Dekany},
  {Dugas}, {Fender}, {Fransson}, {Fremling}, {Goldstein}, {Graham}, {Hale},
  {Horesh}, {Hung}, {Kasliwal}, {Kuin}, {Kulkarni}, {Kupfer}, {Lunnan},
  {Masci}, {Ngeow}, {Nugent}, {Ofek}, {Patterson}, {Petitpas}, {Rusholme},
  {Sai}, {Sfaradi}, {Shupe}, {Sollerman}, {Soumagnac}, {Tachibana}, {Taddia},
  {Walters}, {Wang}, {Yao}, \& {Zhang}}]{Ho+19}
{Ho}, A. Y.~Q., {Goldstein}, D.~A., {Schulze}, S., {et~al.} 2019, \apj, 887,
  169, \dodoi{10.3847/1538-4357/ab55ec}

\bibitem[{{Ho} {et~al.}(2020){Ho}, {Perley}, {Kulkarni}, {Dong}, {De},
  {Chandra}, {Andreoni}, {Bellm}, {Burdge}, {Coughlin}, {Dekany}, {Feeney},
  {Frederiks}, {Fremling}, {Golkhou}, {Graham}, {Hale}, {Helou}, {Horesh},
  {Kasliwal}, {Laher}, {Masci}, {Miller}, {Porter}, {Ridnaia}, {Rusholme},
  {Shupe}, {Soumagnac}, \& {Svinkin}}]{Ho+20}
{Ho}, A. Y.~Q., {Perley}, D.~A., {Kulkarni}, S.~R., {et~al.} 2020, \apj, 895,
  49, \dodoi{10.3847/1538-4357/ab8bcf}

\bibitem[{{Ho} {et~al.}(2021){Ho}, {Margalit}, {Bremer}, {Perley}, {Yao},
  {Dobie}, {Kaplan}, {O'Brien}, {Petitpas}, \& {Zic}}]{Ho+21b}
{Ho}, A. Y.~Q., {Margalit}, B., {Bremer}, M., {et~al.} 2021, arXiv e-prints,
  arXiv:2110.05490.
\newblock \doarXiv{2110.05490}

\bibitem[{{Horesh} {et~al.}(2016){Horesh}, {Hotokezaka}, {Piran}, {Nakar}, \&
  {Hancock}}]{Horesh+16}
{Horesh}, A., {Hotokezaka}, K., {Piran}, T., {Nakar}, E., \& {Hancock}, P.
  2016, \apjl, 819, L22, \dodoi{10.3847/2041-8205/819/2/L22}

\bibitem[{{Huang} {et~al.}(1999){Huang}, {Dai}, \& {Lu}}]{Huang+99}
{Huang}, Y.~F., {Dai}, Z.~G., \& {Lu}, T. 1999, \mnras, 309, 513,
  \dodoi{10.1046/j.1365-8711.1999.02887.x}

\bibitem[{{Katz} {et~al.}(2011){Katz}, {Sapir}, \& {Waxman}}]{Katz+11}
{Katz}, B., {Sapir}, N., \& {Waxman}, E. 2011, arXiv e-prints, arXiv:1106.1898.
\newblock \doarXiv{1106.1898}

\bibitem[{{Kulkarni} {et~al.}(1998){Kulkarni}, {Frail}, {Wieringa}, {Ekers},
  {Sadler}, {Wark}, {Higdon}, {Phinney}, \& {Bloom}}]{Kulkarni+98}
{Kulkarni}, S.~R., {Frail}, D.~A., {Wieringa}, M.~H., {et~al.} 1998, \nat, 395,
  663, \dodoi{10.1038/27139}

\bibitem[{{Kumar} \& {Reville}(2021)}]{Kumar&Reville21}
{Kumar}, N., \& {Reville}, B. 2021, arXiv e-prints, arXiv:2110.09939.
\newblock \doarXiv{2110.09939}

\bibitem[{{Mahadevan} {et~al.}(1996){Mahadevan}, {Narayan}, \&
  {Yi}}]{Mahadevan+96}
{Mahadevan}, R., {Narayan}, R., \& {Yi}, I. 1996, \apj, 465, 327,
  \dodoi{10.1086/177422}

\bibitem[{{Margalit} \& {Metzger}(2019)}]{Margalit&Metzger19}
{Margalit}, B., \& {Metzger}, B.~D. 2019, \apjl, 880, L15,
  \dodoi{10.3847/2041-8213/ab2ae2}

\bibitem[{{Margalit} \& {Piran}(2015)}]{Margalit&Piran15}
{Margalit}, B., \& {Piran}, T. 2015, \mnras, 452, 3419,
  \dodoi{10.1093/mnras/stv1550}

\bibitem[{{Margalit} \& {Piran}(2020)}]{Margalit&Piran20}
---. 2020, \mnras, 495, 4981, \dodoi{10.1093/mnras/staa1486}

\bibitem[{{Margalit} {et~al.}(2021){Margalit}, {Quataert}, \&
  {Ho}}]{Margalit+21}
{Margalit}, B., {Quataert}, E., \& {Ho}, A. Y.~Q. 2021, arXiv e-prints,
  arXiv:2109.09746.
\newblock \doarXiv{2109.09746}

\bibitem[{{Margutti} {et~al.}(2019){Margutti}, {Metzger}, {Chornock}, {Vurm},
  {Roth}, {Grefenstette}, {Savchenko}, {Cartier}, {Steiner}, {Terreran},
  {Margalit}, {Migliori}, {Milisavljevic}, {Alexand er}, {Bietenholz},
  {Blanchard}, {Bozzo}, {Brethauer}, {Chilingarian}, {Coppejans}, {Ducci},
  {Ferrigno}, {Fong}, {G{\"o}tz}, {Guidorzi}, {Hajela}, {Hurley}, {Kuulkers},
  {Laurent}, {Mereghetti}, {Nicholl}, {Patnaude}, {Ubertini}, {Banovetz},
  {Bartel}, {Berger}, {Coughlin}, {Eftekhari}, {Frederiks}, {Kozlova},
  {Laskar}, {Svinkin}, {Drout}, {MacFadyen}, \& {Paterson}}]{Margutti+19}
{Margutti}, R., {Metzger}, B.~D., {Chornock}, R., {et~al.} 2019, \apj, 872, 18,
  \dodoi{10.3847/1538-4357/aafa01}

\bibitem[{{Metzger} \& {Bower}(2014)}]{Metzger&Bower14}
{Metzger}, B.~D., \& {Bower}, G.~C. 2014, \mnras, 437, 1821,
  \dodoi{10.1093/mnras/stt2010}

\bibitem[{{Nakar} \& {Piran}(2011)}]{Nakar&Piran11}
{Nakar}, E., \& {Piran}, T. 2011, \nat, 478, 82, \dodoi{10.1038/nature10365}

\bibitem[{{{\"O}zel} {et~al.}(2000){{\"O}zel}, {Psaltis}, \&
  {Narayan}}]{Ozel+00}
{{\"O}zel}, F., {Psaltis}, D., \& {Narayan}, R. 2000, \apj, 541, 234,
  \dodoi{10.1086/309396}

\bibitem[{{Park} {et~al.}(2015){Park}, {Caprioli}, \& {Spitkovsky}}]{Park+15}
{Park}, J., {Caprioli}, D., \& {Spitkovsky}, A. 2015, \prl, 114, 085003,
  \dodoi{10.1103/PhysRevLett.114.085003}

\bibitem[{{Pe'er}(2012)}]{Peer12}
{Pe'er}, A. 2012, \apjl, 752, L8, \dodoi{10.1088/2041-8205/752/1/L8}

\bibitem[{{Perley} {et~al.}(2021){Perley}, {Ho}, {Yao}, {Fremling}, {Anderson},
  {Schulze}, {Kumar}, {Anupama}, {Barway}, {Bellm}, {Bhalerao}, {Chen}, {Duev},
  {Galbany}, {Graham}, {Gromadzki}, {Guti{\'e}rrez}, {Ihanec}, {Inserra},
  {Kasliwal}, {Kool}, {Kulkarni}, {Laher}, {Masci}, {Neill}, {Nicholl},
  {Pursiainen}, {van Roestel}, {Sharma}, {Sollerman}, {Walters}, \&
  {Wiseman}}]{Perley+21}
{Perley}, D.~A., {Ho}, A. Y.~Q., {Yao}, Y., {et~al.} 2021, \mnras,
  \dodoi{10.1093/mnras/stab2785}

\bibitem[{{Petrosian}(1981)}]{Petrosian81}
{Petrosian}, V. 1981, \apj, 251, 727, \dodoi{10.1086/159517}

\bibitem[{{Prentice} {et~al.}(2018){Prentice}, {Maguire}, {Smartt}, {Magee},
  {Schady}, {Sim}, {Chen}, {Clark}, {Colin}, {Fulton}, {McBrien}, {O'Neill},
  {Smith}, {Ashall}, {Chambers}, {Denneau}, {Flewelling}, {Heinze}, {Holoien},
  {Huber}, {Kochanek}, {Mazzali}, {Prieto}, {Rest}, {Shappee}, {Stalder},
  {Stanek}, {Stritzinger}, {Thompson}, \& {Tonry}}]{Prentice+18}
{Prentice}, S.~J., {Maguire}, K., {Smartt}, S.~J., {et~al.} 2018, \apjl, 865,
  L3, \dodoi{10.3847/2041-8213/aadd90}

\bibitem[{{Ressler} \& {Laskar}(2017)}]{Ressler&Laskar17}
{Ressler}, S.~M., \& {Laskar}, T. 2017, \apj, 845, 150,
  \dodoi{10.3847/1538-4357/aa8268}

\bibitem[{{Rybicki} \& {Lightman}(1979)}]{Rybicki&Lightman79}
{Rybicki}, G.~B., \& {Lightman}, A.~P. 1979, {Radiative processes in
  astrophysics}

\bibitem[{{Sari} {et~al.}(1998){Sari}, {Piran}, \& {Narayan}}]{Sari+98}
{Sari}, R., {Piran}, T., \& {Narayan}, R. 1998, \apjl, 497, L17,
  \dodoi{10.1086/311269}

\bibitem[{{Schroeder} {et~al.}(2020){Schroeder}, {Margalit}, {Fong}, {Metzger},
  {Williams}, {Paterson}, {Alexander}, {Laskar}, {Goyal}, \&
  {Berger}}]{Schroeder+20}
{Schroeder}, G., {Margalit}, B., {Fong}, W.-f., {et~al.} 2020, \apj, 902, 82,
  \dodoi{10.3847/1538-4357/abb407}

\bibitem[{{Sironi} \& {Spitkovsky}(2009)}]{Sironi&Spitkovsky09}
{Sironi}, L., \& {Spitkovsky}, A. 2009, \apj, 698, 1523,
  \dodoi{10.1088/0004-637X/698/2/1523}

\bibitem[{{Sironi} \& {Spitkovsky}(2011)}]{Sironi&Spitkovsky11}
---. 2011, \apj, 726, 75, \dodoi{10.1088/0004-637X/726/2/75}

\bibitem[{{Spitkovsky}(2008)}]{Spitkovsky08}
{Spitkovsky}, A. 2008, \apjl, 682, L5, \dodoi{10.1086/590248}

\bibitem[{{Spitzer}(1956)}]{Spitzer56}
{Spitzer}, L. 1956, {Physics of Fully Ionized Gases}

\bibitem[{{Stepney}(1983)}]{Stepney83}
{Stepney}, S. 1983, \mnras, 202, 467, \dodoi{10.1093/mnras/202.2.467}

\bibitem[{{Tan} {et~al.}(2001){Tan}, {Matzner}, \& {McKee}}]{Tan+01}
{Tan}, J.~C., {Matzner}, C.~D., \& {McKee}, C.~F. 2001, \apj, 551, 946,
  \dodoi{10.1086/320245}

\bibitem[{{Tran} \& {Sironi}(2020)}]{Tran&Sironi20}
{Tran}, A., \& {Sironi}, L. 2020, \apjl, 900, L36,
  \dodoi{10.3847/2041-8213/abb19c}

\bibitem[{{Warren} {et~al.}(2018){Warren}, {Barkov}, {Ito}, {Nagataki}, \&
  {Laskar}}]{Warren+18}
{Warren}, D.~C., {Barkov}, M.~V., {Ito}, H., {Nagataki}, S., \& {Laskar}, T.
  2018, \mnras, 480, 4060, \dodoi{10.1093/mnras/sty2138}

\bibitem[{{Weiler} {et~al.}(2002){Weiler}, {Panagia}, {Montes}, \&
  {Sramek}}]{Weiler+02}
{Weiler}, K.~W., {Panagia}, N., {Montes}, M.~J., \& {Sramek}, R.~A. 2002,
  \araa, 40, 387, \dodoi{10.1146/annurev.astro.40.060401.093744}

\bibitem[{{Xu} {et~al.}(2020){Xu}, {Spitkovsky}, \& {Caprioli}}]{Xu+20}
{Xu}, R., {Spitkovsky}, A., \& {Caprioli}, D. 2020, \apjl, 897, L41,
  \dodoi{10.3847/2041-8213/aba11e}

\bibitem[{{Zauderer} {et~al.}(2011){Zauderer}, {Berger}, {Soderberg}, {Loeb},
  {Narayan}, {Frail}, {Petitpas}, {Brunthaler}, {Chornock}, {Carpenter},
  {Pooley}, {Mooley}, {Kulkarni}, {Margutti}, {Fox}, {Nakar}, {Patel},
  {Volgenau}, {Culverhouse}, {Bietenholz}, {Rupen}, {Max-Moerbeck}, {Readhead},
  {Richards}, {Shepherd}, {Storm}, \& {Hull}}]{Zauderer+11}
{Zauderer}, B.~A., {Berger}, E., {Soderberg}, A.~M., {et~al.} 2011, \nat, 476,
  425, \dodoi{10.1038/nature10366}

\end{thebibliography}
\bibliographystyle{aasjournal}



\end{document}